\documentclass[aps,prd,superscriptaddress,showkeys,showpacs,twocolumn,a4paper]{revtex4-1}
\usepackage{ulem}
\usepackage{color}
\usepackage{graphicx}
\usepackage{soul}
\usepackage{appendix}
\usepackage{epsfig}
\usepackage[usenames,dvipsnames,svgnames]{xcolor}
\definecolor{purple}{rgb}{0.3,0,0.9} 
\definecolor{darkteal}{HTML}{045D5D}
\usepackage[colorlinks=true,linkcolor= cyan, citecolor = blue, urlcolor = cyan ]{hyperref}

\usepackage{xurl}                    
\hypersetup{breaklinks=true}         
\usepackage{epstopdf}
\usepackage{amsmath,amssymb,amsthm,amsfonts} 
\usepackage{subfig}
\usepackage{comment}
\usepackage{float}

\usepackage{float}

\newcommand{\be}{\begin{equation}}
\newcommand{\ee}{\end{equation}}
\newcommand{\ba}{\begin{eqnarray}}
\newcommand{\ea}{\end{eqnarray}}
\newcommand{\nn}{\nonumber\\}

\begin{document}
\title{Melting of heavy quarkonia in QGP using deep neural networks}
  \author{Mohammad Yousuf Jamal}
\email{yousufjml5@gmail.com}
 \affiliation{Key Laboratory of Quark and Lepton Physics (MOE) \& Institute of Particle Physics, Central China Normal University, Wuhan 430079, China}

\author{Fu-Peng Li}
\email[]{fpli@fudan.edu.cn}
\affiliation{Key Laboratory of Quark and Lepton Physics (MOE) \& Institute of Particle Physics, Central China Normal University, Wuhan 430079, China}
\affiliation{Artificial Intelligence and Computational Physics Research Center, Central China Normal University, Wuhan 430079, China}
\affiliation{Key Laboratory of Nuclear Physics and Ion-beam Application (MOE) \& Institute of Modern Physics, Fudan University, Shanghai 200433, China}
\affiliation{Shanghai Research Center for Theoretical Nuclear Physics,
NSFC and Fudan University, Shanghai 200438, China}

\author{Long-Gang Pang}
\email[]{lgpang@ccnu.edu.cn}
\affiliation{Key Laboratory of Quark and Lepton Physics (MOE) \& Institute of Particle Physics, Central China Normal University, Wuhan 430079, China}
\affiliation{Artificial Intelligence and Computational Physics Research Center, Central China Normal University, Wuhan 430079, China}

\author{Guang-You Qin}
\email[]{guangyou.qin@ccnu.edu.cn}
\affiliation{Key Laboratory of Quark and Lepton Physics (MOE) \& Institute of Particle Physics, Central China Normal University, Wuhan 430079, China}
\affiliation{Artificial Intelligence and Computational Physics Research Center, Central China Normal University, Wuhan 430079, China}

\begin{abstract}
Machine learning techniques have emerged as powerful tools for tackling non-perturbative challenges in quantum chromodynamics. 
In this study, we introduce a data-driven framework employing deep neural networks to systematically predict the temperature-dependent behavior of the screening mass $m_D(T)$ and the strong coupling constant $\alpha_s(T)$ within a quark-gluon plasma medium. These medium-sensitive quantities are subsequently employed to compute the thermal widths $\Gamma_{\text{n}}(T)$ and binding energies $E_B(T)$ of heavy quarkonia states, specifically charmonia and bottomonia, by numerically solving the Schr\"odinger equation with medium-modified heavy quark potentials. To estimate the dissociation temperatures $T_d$ of various quarkonia states, we employ two complementary dissociation criteria: the conventional one, where $2E_B(T_d) = \Gamma_{\text{n}}(T_d)$, and an additional lower bound criterion defined by $E_B(T_d) = 3T_d$. This dual-criterion approach provides a more constrained and physically motivated estimate of the temperature range over which quarkonia states dissolve in the QGP environment. Our machine learning-enhanced predictions show excellent agreement with available lattice QCD results, especially for the ground states $\Upsilon(1S)$ and $J/\psi$, and offer new perspectives on the sequential suppression pattern detected in relativistic heavy-ion collision experiments. Overall, this work advances the quantitative description of quarkonium suppression and demonstrates the prospect of modern machine learning methods to bridge theoretical predictions and experimental observations, thereby contributing significantly to QGP tomography.
\end{abstract}
\maketitle

\section{Introduction}

Ultra-relativistic heavy-ion collisions at RHIC and the LHC have established that the quark-gluon plasma (QGP) behaves as a nearly perfect fluid, exhibiting strong collective flow and an exceptionally low shear viscosity to entropy density ratio~\cite{Heinz:2004qz}. This remarkable finding stands in stark contrast to early theoretical expectations of a weakly interacting gas of quarks and gluons, 
emphasizing the necessity for precision probes to unravel the emergent properties of this strongly coupled medium. Heavy quarkonia ($Q\bar{Q}$ bound states) have proven to be particularly sensitive indicators of deconfinement and the QGP's screening behavior~\cite{Matsui:1986dk, McLerran:1986zb, Back:2004je, Nilima:2024nvd}, with their suppression patterns encoding information on color screening, parton density, and medium-induced dissociation mechanisms~\cite{Chu:1988wh, Koike:1991mf}.

Being color-singlet states composed of heavy quark-antiquark pairs ($c\bar{c}$, $b\bar{b}$), quarkonia occupy a unique position in QCD phenomenology~\cite{Burnier:2009yu, Dumitru:2009fy,Pooja:2024rnn}. Their production occurs predominantly in the initial hard scatterings of a collision, allowing them to probe the earliest stages of the medium before significant collective evolution ensues. The interplay of perturbative and non-perturbative scales in QCD, ranging from the large heavy quark mass to confinement-driven string dynamics~\cite{Voloshin:2007dx, Brambilla:2010cs, Patrignani:2012an}, has inspired a diverse array of models for quarkonium production and suppression, including the color evaporation model~\cite{Frawley:2008kk, Amundson:1996qr}, color singlet and octet mechanisms~\cite{Berger:1980ni, Adamczyk:2012ey}, and recombination scenarios~\cite{Silvestre:2008tw,Singh:2023zxu}.

The foundational idea, proposed by Matsui and Satz~\cite{Matsui:1986dk}, states that color screening in the QGP weakens the confining force between the heavy quark and antiquark, resulting in quarkonium dissociation. This picture has evolved through refinements incorporating thermal broadening, dynamical screening, and real-time effects~\cite{Mocsy:2004bv, Agotiya:2008ie, Sebastian:2022sga}. In practice, medium modifications to the heavy quark potential, often formulated as a screened Cornell-type potential~\cite{Eichten:1978tg, Eichten:1979ms, Chung:2008sm, Nilima:2024qzx}, play a central role, modified via dielectric permittivity using the hard thermal loop approximation~\cite{Laine:2006ns, Strickland:2011aa, Margotta:2011ta, Thakur:2013nia}. This gives us a medium modified complex potential between $Q\bar{Q}$ bound states. The real part governs the temperature-dependent binding energy $E_B(T)$, while the imaginary part yields the thermal decay width $\Gamma_{\text{n}}(T)$, 

In the standard approach, quarkonium dissociation is identified by equating the thermal width to twice the binding energy, {$\Gamma_{\text{n}}(T_d) = 2E_B(T_d)$ where \(T_d\) denotes
the dissociation temperature~\cite{Mocsy:2007jz, Burnier:2009yu}. To further refine this estimate, we additionally impose a physically motivated lower bound criterion: the state is considered too weakly bound to survive when its binding energy drops to the order of the mean thermal energy of the medium, i.e., $E_B(T_d) = 3T_d$~\cite{Digal:2001iu, Agotiya:2008ie}.} The simultaneous use of both criteria constrains the possible dissociation temperature window, yielding a more robust estimate consistent with lattice and spectral function analyses. 

A key innovation in this work is the use of machine learning (ML) to enhance the determination of the Debye screening mass $m_D(T)$ and the strong coupling $\alpha_s(T)$, crucial inputs for the in-medium potential. By training a deep neural network (DNN) on lattice QCD results, we obtain non-perturbative, data-driven estimates of $m_D$ and $\alpha_s$, which in turn lead to improved predictions for $E_B(T)$, $\Gamma_{\text{n}}(T)$, and hence, $T_d$ for various charmonium and bottomonium states. This ML-augmented framework bridges the rigor of first-principles calculations with the computational efficiency of modern data science tools, providing a powerful approach for QGP tomography.

The remainder of this paper is organized as follows. Section \eqref{sec:HQP} outlines the theoretical framework, including the ML-based extraction of the Debye mass and the medium-modified potential. Section \eqref{sec:results} presents our results for the dissociation temperatures using both dissociation criteria, and compares them with lattice QCD and some other available data. Section \eqref{sec:Con} summarizes our conclusions and provides an outlook for future developments. Some discussions on numerical analysis related to the current work are presented in Appendix \eqref{sec:FDM}.

\section{Formalism}
\label{sec:HQP}
The behavior of heavy quarkonia in a hot QCD medium crucially depends on how the inter-quark potential is modified by color screening and medium-induced dissipation. This section details the theoretical framework employed in this work, including the construction of the in-medium quarkonium potential, the machine-learning-based determination of the Debye mass, and the extraction of binding energies, thermal widths, and dissociation temperatures using both the conventional and lower-bound criteria.

\subsection{Quarkonium Potential in the hot QGP Medium} 
\label{sec:potential}
In vacuum, the static quark-antiquark interaction is well described by the Cornell potential~\cite{Eichten:1978tg, Eichten:1979ms}, which captures both the short-range Coulombic attraction and long-range confining force:
\begin{equation}
V_{\text{vac}}(r) = -\frac{\alpha}{r} + \sigma r,
\label{eq:Vac}
\end{equation}
where $\alpha$ is the effective strong coupling constant and $\sigma$ denotes the string tension. At finite temperature, screening effects in the QGP medium weaken both components of the potential. Following the approach in Refs.~\cite{Agotiya:2008ie, Digal:2001iu}, we introduce medium modifications via the dielectric permittivity $\epsilon(k)$ in momentum space:
\begin{equation}
\tilde{V}(k, T) = \frac{\tilde{V}_{\text{vac}}(k)}{\epsilon(k)}.
\label{eq:Vft}
\end{equation}
The Fourier transform of the vacuum Cornell potential is given by:
\begin{equation}
\tilde{V}_{\text{vac}}(k) = -\sqrt{\frac{2}{\pi}} \left( \frac{\alpha}{k^2} + \frac{2\sigma}{k^4} \right).
\label{eq:Vft1}
\end{equation}
The in-medium potential is obtained by inverse Fourier transform of Eq.\eqref{eq:Vft} where the complex form of $\epsilon(k)$  can be found in ~Ref.\cite{Jamal:2018mog}. Due to the complex form of $\epsilon(k)$, the medium-modified potential also splits into real and imaginary parts. The real potential is obtained as:
\begin{equation}
\text{Re}[V(r,T)] = -\alpha \frac{e^{-m_D r}}{r} + \frac{2\sigma}{m_D} \left(1 - e^{-m_D r}\right) - \sigma r e^{-m_D r}, 
\label{eq:ReV}
\end{equation}
\begin{equation}
\text{Im}[V(r,T)] = \text{Im}V_{1}(r, T) + \text{Im}V_{2}(r, T),
\label{eq:ImV}
\end{equation}
where
\begin{align}
\text{Im}V_{1}(r, T) &= -2 \alpha T \int_{0}^{\infty} \frac{dz}{(z^2 + 1)^2} \left(1 - \frac{\sin (m_D r z)}{ m_D r z}\right),\nn
\text{Im}V_{2}(r, T) &= \frac{4 \sigma T}{m_D^2} \int_{0}^{\infty} \frac{dz}{z(z^2 + 1)^2} \left(1 - \frac{\sin (m_D r z)}{ m_D r z}\right).
\label{eq:im}
\end{align}
The presence of an imaginary part accounts for Landau damping and inelastic scattering processes that broaden the quarkonium state in the medium~\cite{Laine:2006ns, Burnier:2009yu}.

The screening mass \(m_D(T)\) modifies both the short-range Coulomb term and the long-range string term of the potential. Specifically, it introduces an exponential damping that weakens the attractive force at large distances, reflecting color screening in the deconfined medium. The Coulombic term becomes Yukawa-like, while the linear confining term is partially screened and suppressed.

It is important to note that the Debye screening mass  $m_D(T)$ is not just an algebraic function of the quasi-parton masses, but is more deeply rooted in the underlying thermal structure of the medium. In hot QCD, the Debye mass originates from the static limit of the longitudinal gluon self-energy and is fundamentally expressed in terms of the thermal distribution functions of quarks and gluons as:
\begin{align}
m_D^2 = -4 \pi \alpha_{s}(T) \bigg(2 N_c \int \frac{d^3 p}{(2 \pi)^3} \partial_p f_g ({\bf p}) &\nn
+ N_f \int \frac{d^3 p}{(2 \pi)^3} \partial_p( f_q ({\bf p})+f_{\bar q} ({\bf p}))\bigg).
\label{dm}    
 \end{align}

In quasi-particle models, these distribution functions  $f_{g,q,\bar{q}}(p)$ are modified to include medium effects, typically by introducing temperature-dependent effective masses $ m_{g/q}(T)$ in the dispersion relation $E(p) = \sqrt{p^2 + m^2(T)}$ or by adding fugacity-like corrections. As a result, the Debye mass becomes implicitly dependent on the quasi-particle properties through these modified distributions. While our approach computes  $m_D(T)$ using a formula derived from leading-order hard thermal loop (HTL) theory, the inputs, specifically the quasi-parton masses obtained via machine learning, {already encode non-trivial thermal corrections \cite{Song:2015ykw,Sambataro:2024mkr,Soloveva:2023tvj}.} Therefore, the Debye mass in this framework effectively captures the collective response of the QGP medium and remains tightly connected to the quasi-particle description.

\subsection{Machine Learning Determination of Debye Mass and Strong Coupling}
\label{sec:ML}
A crucial input parameter for the medium-modified potential is the Debye mass \(m_D(T)\), which incorporates non-perturbative effects beyond leading-order HTL calculations~\cite{Levai:1997yx}. In this work, we utilize a Deep-Learning Quasi-Parton gas Model (DLQPM) to extract \(m_D(T)\) and the running coupling constant \(\alpha_s(T)\) directly from lattice QCD data~\cite{Li:2022ozl, Li:2025csc, HotQCD:2014kol}. The training methodology comprises the following steps:

\begin{figure}[ht!]
    \centering
    \includegraphics[width=0.48\textwidth]{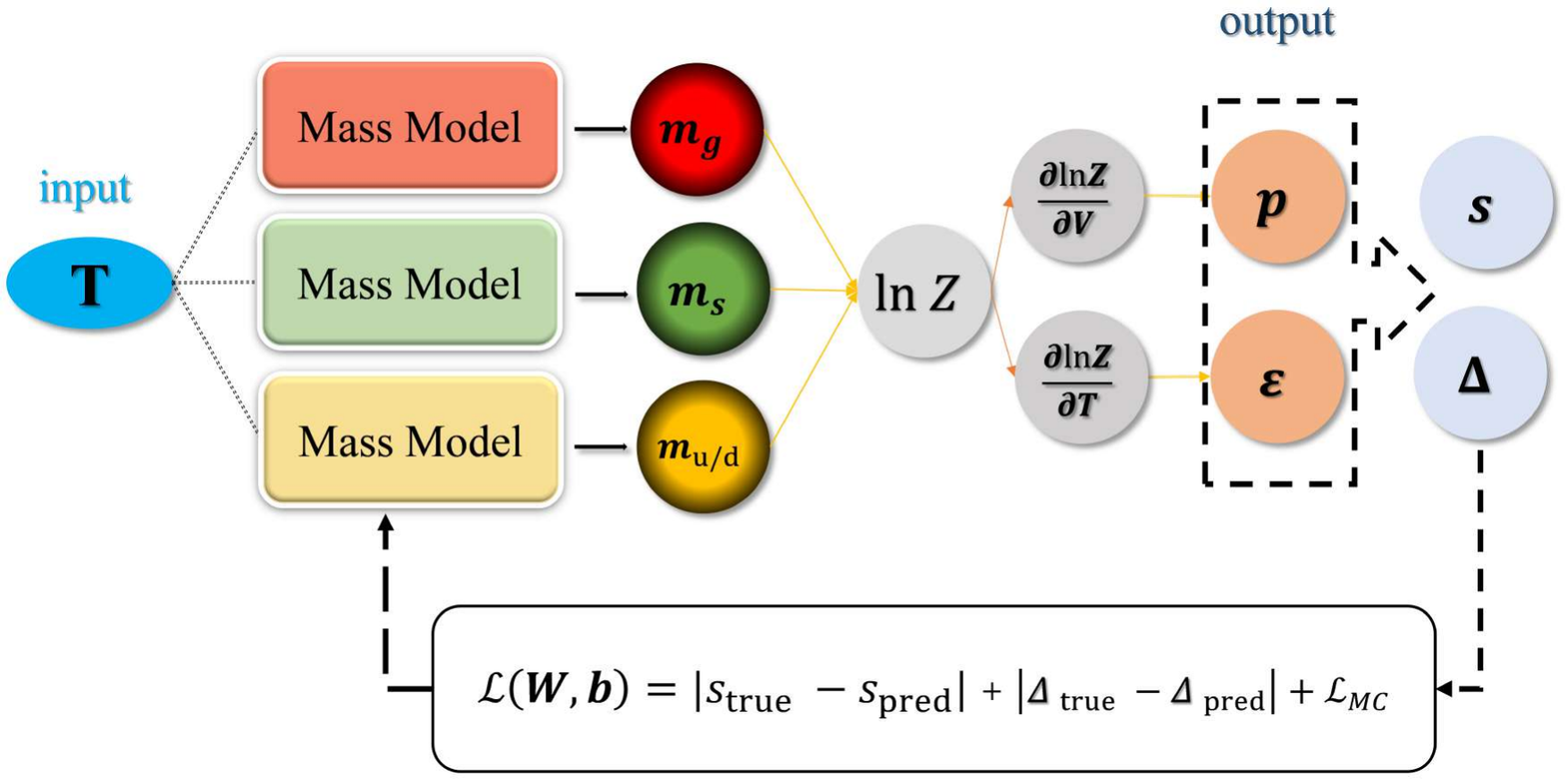}
    \caption{Sketch of the DNN used to extract the temperature-dependent quasi-parton masses \(m_{u/d}(T)\), \(m_{s}(T)\), and \(m_{g}(T)\)~\cite{Li:2022ozl, Li:2025csc}. {The symbols \(W\) and \(b\) denote the trainable weights and biases of the neural network}.}
    \label{fig:DNN}
\end{figure}

\begin{itemize}
    \item \textbf{Architecture:} The framework of DLQPM, sketched in Fig.~\ref{fig:DNN}, is designed to reconstruct the QCD equation of state (EoS) at zero chemical potential. The model incorporates three mass subnetworks for light \(u/d\) quarks, strange quarks, and gluons, respectively. Each subnetwork is implemented as an eight-layer Residual Neural Network (ResNet) with swish activation functions and 32 neurons per layer.

    \item \textbf{Training set:} We employ 50 numerical data points from HotQCD calculations~\cite{HotQCD:2014kol}, covering the temperature range \( T/T_c \in [1, 3] \), where the critical temperature is taken as \(T_c = 0.15~\mathrm{GeV}\).

    \item \textbf{Objective:} Given the quasi-parton masses, the EoS is obtained from the partition function. {The network parameters (weights and biases) are optimized by minimizing the mean squared error between the model predictions and lattice QCD results for the entropy density \(s(T)\) and the trace anomaly \(\Delta(T) = \varepsilon(T) - 3p(T)\). In addition, we impose high-temperature constraints inspired by HTL calculations through the loss term \(\mathcal{L}_{MC}\), which is activated at \(T > 2.5\,T_{\rm cut}\),}
    \begin{align}
    \mathcal{L}_{MC} &= \left| R_{g/q} - \frac{3}{2} \right| 
    +  \left| \frac{m_s - m_{u/d}}{\overline{m}_s - \overline{m}_{u/d}} - 1 \right| , \nonumber \\
    R_{g/q} &= \frac{m_{g,T>2.5T_{\rm cut}}}{m_{u/d,T>2.5T_{\rm cut}}}
    = \sqrt{\frac{3}{2}\left(\frac{N_c}{3} + \frac{N_f}{6}\right)},
    \label{eq:MCC}
    \end{align}
    where \(m_{u/d}\), \(m_s\), and \(m_g\) are the effective masses output by the DLQPM, and \(\overline{m}_{u/d}\), \(\overline{m}_s\) denote the{color{red} corresponding current masses (\(\overline{m}_s = 95~\mathrm{MeV}\))}. We set \(T_{\rm cut} = T_c\), and the choice \(T > 2.5\,T_{\rm cut}\) defines a high-temperature region where HTL expectations are reliable and can be used to constrain the asymptotic behaviour of the masses. After \(5\times 10^{4}\) training epochs, the total loss converges to values of order \(10^{-5}\).

    \item \(m_D(T)\) \textbf{and} \(\alpha_s(T)\): Once the quasi-parton masses are obtained using the DLQPM, the strong coupling constant \(\alpha_s(T)\) is determined from the thermal gluon and light-quark masses~\cite{Plumari:2011mk}:
    \begin{align}
        m_g^2(T) &= \frac{1}{6}g_s^2(T)\bigg[ \Big(N_c + \frac{1}{2}N_f \Big)T^2 \bigg], \nonumber \\
        m_{u,d}^2(T) &= \frac{N_c^2 - 1}{8N_c}g_s^2(T) \, T^2,
    \end{align}
    leading to
    \begin{equation}
        g_s^2(T) = \frac{m_g^2(T) + m_{u,d}^2(T)}
        {\frac{1}{6}\Big[ \Big(N_c + \frac{1}{2}N_f \Big)T^2 \Big] + \frac{N_c^2 - 1}{8N_c}T^2},
        \label{eq:g2}
    \end{equation}
    and \(\alpha_s(T) = g_s^2(T) / (4\pi)\).
\end{itemize}

The relations above connect the quasi-parton thermal masses to the temperature-dependent strong coupling \(\alpha_s(T)\), which in turn provides a key input for the Debye screening mass. Physically, the Debye mass \(m_D(T)\) increases with temperature as a consequence of the growing density of color charges in the deconfined medium. This enhancement reflects the QGP’s increasing ability to
screen long-range color interactions, thereby weakening the confining potential between quarks and antiquarks at elevated temperatures. Now using these inputs, $m_D(T)$ can be computed via the following gauge-invariant relation: 
\begin{equation}
m_D(T) = T \sqrt{4\pi\alpha_s(T) \left( \frac{N_c}{3} + \frac{N_f}{6} \right)} .
\label{eq:md}
\end{equation}

{To quantify the uncertainty associated with the DLQPM extraction, we retrain the model ten times, obtaining ten distinct sets of quasi-parton masses. From these, we compute the mean and variance of \(m_D(T)\) and \(\alpha_s(T)\), and propagate them through the potential to the binding energies and thermal widths. The resulting error bands shown in the figures of Sec.~\ref{sec:results} therefore reflect the spread over these independent trainings. Additional training diagnostics, including the evolution of the total loss and learning rate with the epoch, are presented in Appendix~\ref{sec:DNN}.}

\begin{figure*}[ht!]
    \centering
    \includegraphics[width=0.48\textwidth]{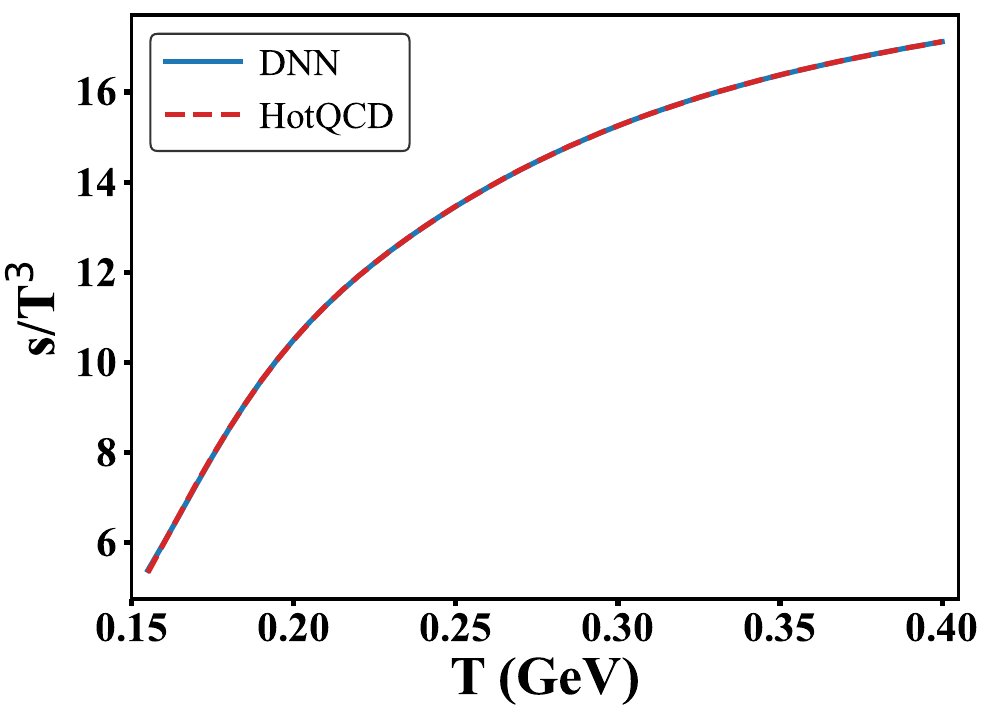}
    \includegraphics[width=0.48\textwidth]{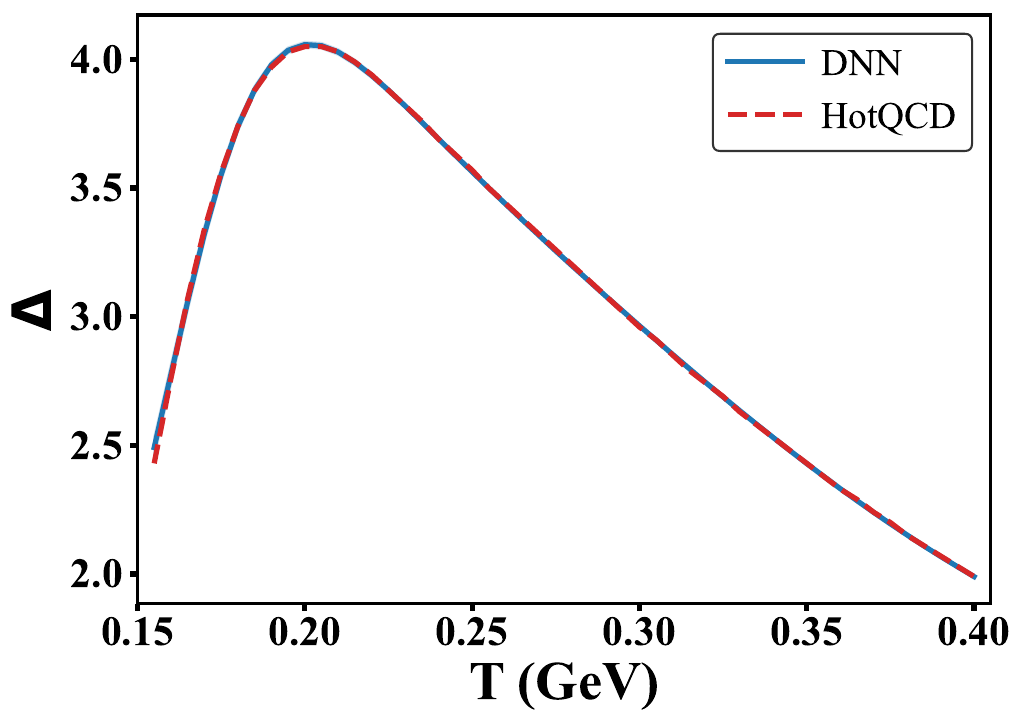}
   \caption{Entropy density \(s/T^{3}\) (left) and trace anomaly \(\Delta = \varepsilon - 3p\) (right) as functions of temperature, comparing the DLQPM reconstruction with HotQCD lattice results~\cite{HotQCD:2014kol}.}
    \label{fig:TH}
\end{figure*}

\begin{figure}[ht!]
    \centering
    \includegraphics[width=1.0\linewidth]{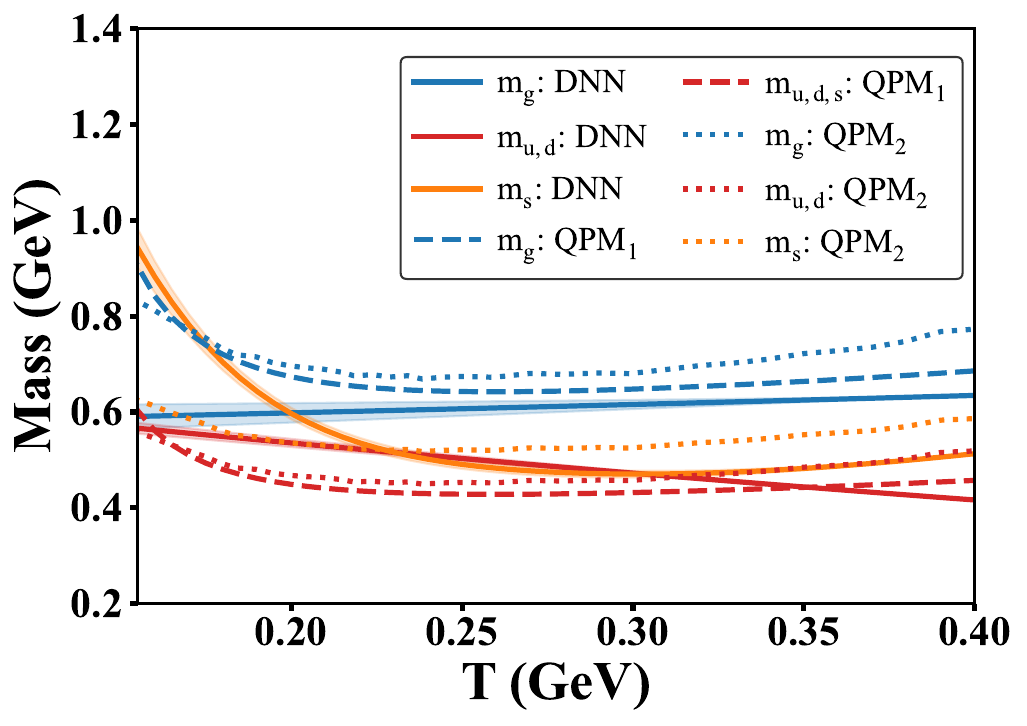}
\caption{Temperature dependence of the effective quasi-parton masses for light quarks, strange quarks, and gluons obtained from the present DLQPM, compared with representative quasi-particle model results from Refs.~\cite{Liu:2021dpm,Mykhaylova:2019wci}.}
    \label{fig:masses}
\end{figure}

\begin{figure}[ht!]
    \centering
    \includegraphics[width=1.0\linewidth]{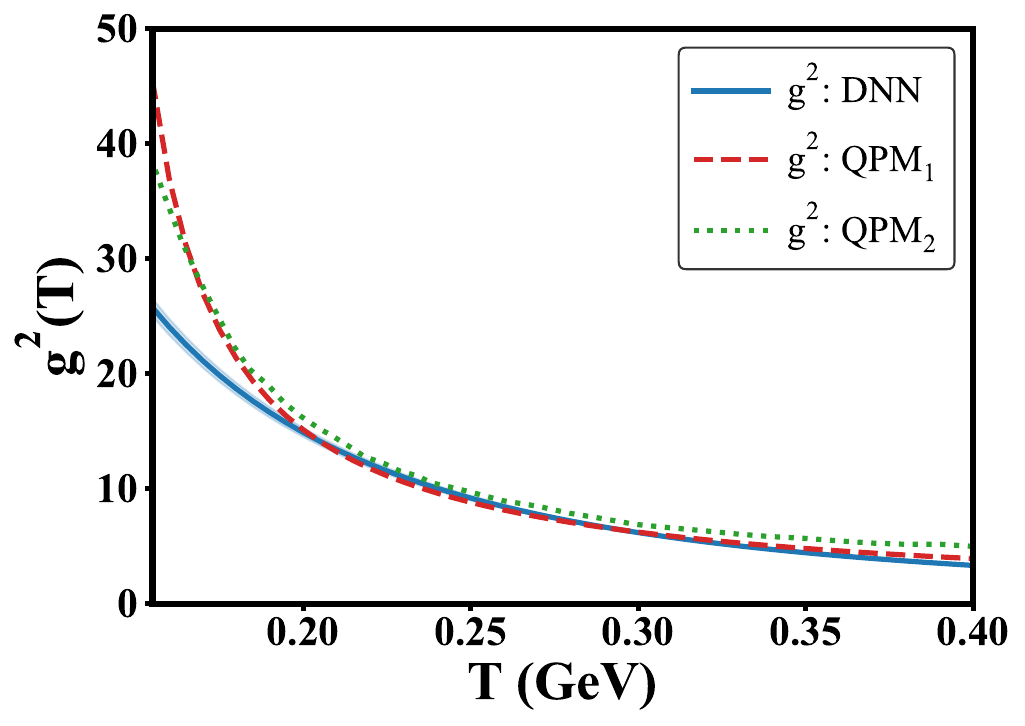}
\caption{Temperature dependence of the strong coupling \(g^{2}(T)\) extracted from the present DLQPM, compared with representative quasi-particle model parametrisations from Refs.~\cite{Liu:2021dpm,Mykhaylova:2019wci}.}
    \label{fig:g2}
\end{figure}

{As a validation of the DLQPM, in Fig.~\ref{fig:TH} we compare the entropy density \(s/T^{3}\) and the trace anomaly \(\Delta(T)\) obtained from the network with the HotQCD data~\cite{HotQCD:2014kol}. The excellent agreement over the range \(T/T_{c}\in[1,3]\) demonstrates that the learned quasi-parton masses reproduce the lattice equation of state. In Fig.~\ref{fig:masses}  we show the resulting effective masses \(m_{u/d}(T)\), \(m_{s}(T)\), and \(m_{g}(T)\) and confront them with representative quasi-particle model parametrisations~\cite{Liu:2021dpm,Mykhaylova:2019wci}. The overall consistency, including the expected hierarchy between light and strange quarks, suggests that the DLQPM effectively captures the main medium effects encoded in these models. The corresponding strong coupling \(g^{2}(T)\), reconstructed from Eq.~\eqref{eq:g2}, is displayed in Fig.~\ref{fig:g2}  and compared with standard parametrisations; the modest differences at intermediate temperatures can be traced back to distinct implementation choices in the underlying quasi-particle fits.}

While our framework is data-driven and anchored to lattice QCD results via ML training, the analytic forms used to extract $\alpha_s(T)$ and $m_D(T)$ incorporate leading-order HTL structures. Thus, the approach captures non-perturbative trends through ML optimization but still relies on perturbative-inspired expressions. Therefore, we interpret the non-perturbative nature of the model in a practical, hybrid sense—combining perturbative inputs with lattice-constrained ML outputs to enhance the authenticity of the potential. This ML-derived Debye mass exhibits a smooth transition to lattice QCD predictions, thereby ensuring that the medium screening and damping effects are fundamentally grounded in non-perturbative QCD data rather than being dependent solely on perturbative approximations~\cite{Digal:2001iu, Agotiya:2008ie}. In particular, this data-driven approach ensures that the potential's temperature dependence is anchored to first-principles lattice QCD results, especially in the crossover region ($1.0 \leq T/T_c \leq 1.5$), where conventional perturbative methods are known to break down~\cite{Mocsy:2007jz,Agotiya:2008ie}. The resulting potential then serves as input for the Schr\"odinger equation (Sec.~\ref{sec:Schrodinger}) to compute binding energies and widths.

\subsection{Schr\"odinger Equation and Numerical Solutions} 
\label{sec:Schrodinger}
The temperature-dependent binding energies $E_B(T)$ and radial wavefunctions $\psi_n(r)$ are obtained by solving the time-independent radial Schr\"odinger equation using the real part of the in-medium heavy-quark potential:
\begin{align}
\left[ -\frac{1}{2\mu_{Q\bar{Q}}} \frac{d^2}{dr^2} + \frac{\ell(\ell+1)}{2\mu_{Q\bar{Q}} r^2} + \text{Re}[V(r,T)] \right] \psi_n(r) &\nn
= E_n(T) \psi_n(r),    
\end{align}
where $\mu_{Q\bar{Q}} = m_Q/2$ is the reduced mass of the heavy quark-antiquark pair and $\ell$ is the orbital angular momentum quantum number (taken  $\ell=0$ for the $S$-wave ground and first excited states). To accurately resolve the delicate balance between short-range Coulomb attraction and long-range screening, we employ a finite difference method (FDM) with second-order central difference stencils and Neumann boundary conditions at $r=r_{\text{min}}$ and $r=r_{\text{max}}$. The radial coordinate is discretized into $N=4000$ grid points extending up to $r_{\text{max}} = 30$~fm to ensure convergence even at low binding energies near the dissociation point. The resulting large sparse matrix eigenvalue problem is solved iteratively for each temperature step using a combination of bisection and inverse iteration, allowing us to track both the ground and excited states continuously as functions of $T/T_c \in [1, 3]$. This approach ensures robust extraction of weakly bound states near the deconfinement crossover, where the binding energies become comparable to the thermal energy scale. We have included Appendix~\ref{sec:FDM}, which provides a detailed introduction to the finite difference method (FDM) and its comparison with alternative numerical approaches.

\subsection{Thermal Width} 
\label{sec:width}
The thermal width \(\Gamma_n(T)\) characterizes the in-medium decay rate, or equivalently the inverse lifetime, of a quarkonium state embedded in the QGP. It provides a quantitative measure of the extent to which a bound state becomes unstable due to interactions with the thermal medium. Physically, this width reflects the spectral broadening of quarkonium states arising from medium-induced processes such as dissociation and scattering with thermal partons.

In the potential model framework, the interaction between a heavy quark and antiquark is described by a complex potential, where the real part governs the binding and the imaginary part encodes dissipative effects due to the QGP. The imaginary component acts as a perturbative contribution to the otherwise real Cornell-type potential. Within the formalism of first-order quantum mechanical perturbation theory, this imaginary term induces a finite decay width for the bound state~\cite{Laine:2006ns, Brambilla:2008cx}. The thermal width is thus computed as the expectation value of the imaginary part of the potential over the normalized quarkonium wavefunction:

\begin{equation}
\Gamma_n(T) = - \int_0^\infty 4\pi r^2\, |\psi_n(r)|^2\, \text{Im}[V(r,T)]\, dr.
\label{eq:gamma}
\end{equation}
This formulation captures how the imaginary part of the in-medium potential introduces a thermal decay channel for the heavy quark-antiquark pair. The imaginary potential arises from processes like Landau damping and gluodissociation, which represent the scattering of the quarkonium with the thermal gluons and quarks of the surrounding plasma. These interactions lead to decoherence and eventual dissociation of the bound state, manifesting as a broadening of its spectral peak. Here, \(\psi_n(r)\) is the radial wavefunction corresponding to the \(n^\text{th}\) quarkonium state, obtained by solving the Schr\"odinger equation using the real part of the temperature-dependent complex potential. The term \(\text{Im}[V(r,T)]\), provided explicitly in Eq.~\eqref{eq:im}, encapsulates the dissipative features of the QGP medium. The negative sign in Eq.\eqref{eq:gamma} ensures that a negative imaginary potential translates to a positive decay width, consistent with physical intuition.

The magnitude of \(\Gamma_n(T)\) governs how rapidly a quarkonium state is likely to decay or dissolve under thermal fluctuations. A larger thermal width corresponds to a shorter survival time, making the bound state more susceptible to melting. As the temperature of the QGP increases, both the screening of the color potential and the inelastic damping effects become more pronounced, enhancing \(\text{Im}[V(r,T)]\) and thereby increasing \(\Gamma_n(T)\). When the thermal width becomes comparable to or exceeds the binding energy, the bound state is effectively destroyed. In this study, we systematically evaluate \(\Gamma_n(T)\) for both ground and excited states of charmonium and bottomonium. These widths, in conjunction with the corresponding binding energies, are used to determine the dissociation temperatures of the quarkonium states based on well-motivated physical criteria.

\subsection{Dissociation Criteria} 
\label{sec:dissociation}
A precise determination of the dissociation temperature $T_d$ for quarkonium states is crucial for interpreting their sequential suppression patterns in heavy-ion collisions~\cite{Matsui:1986dk, Digal:2001iu}. In this study, we adopt two well-motivated and complementary criteria to estimate the temperature range over which a given quarkonium state melts in the quark-gluon plasma. The first is the conventional thermal width criterion, which asserts that a resonance ceases to exist as a distinguishable bound state once its thermal decay width becomes comparable to its binding energy~\cite{Mocsy:2007jz, Burnier:2009yu, Laine:2006ns}. Following standard practice, we employ the quantitative condition $\Gamma_n(T_d) = 2E_B(T_d)$. Physically, this implies that the mean time for thermal dissociation due to Landau damping and gluodissociation is on the order of the inverse binding time, leading to significant broadening of the quarkonium spectral function and the disappearance of a well-defined peak~\cite{Brambilla:2008cx}.

However, it is well-known that potential models can underestimate thermal fluctuations, especially near the crossover region where non-perturbative effects persist. To address this, we incorporate an additional lower-bound criterion for dissociation, inspired by the argument in Refs.~\cite{Digal:2001iu, Agotiya:2008ie, Wong:2004zr}. A quarkonium state is considered too weakly bound to survive if its binding energy drops to the scale of the average kinetic energy of thermal partons. For an ultra-relativistic medium, this scale is approximately $3T$, leading to the condition $E_B(T_d) = 3T_d$. This lower bound accounts for the fact that even if the width is small, a shallow binding renders the state highly susceptible to break-up via thermal collisions or screening. { It is important to note that \(E_{B}(T_d) = 3T_d\) corresponds to the average kinetic energy of a thermal excitation in the weak–coupling (Stefan–Boltzmann) limit. In quasi–particle models, the thermal masses tend to reduce the kinetic energy, so this criterion acts as a conservative lower bound.}

Both conditions are solved numerically, applied to the temperature-dependent binding energy and thermal width obtained from the Schr\"odinger solution. This dual-criterion framework yields a physically constrained temperature window for quarkonium dissociation, providing insight into how robustly each state survives at different QGP temperatures. The extracted dissociation temperatures for $J/\psi$, $\psi(2S)$, $\Upsilon(1S)$, and $\Upsilon(2S)$ are presented and compared with lattice QCD and experimental observations in Sec.\eqref{sec:results}.

\begin{figure}[ht!]
    \centering
    \includegraphics[width=0.48\textwidth]{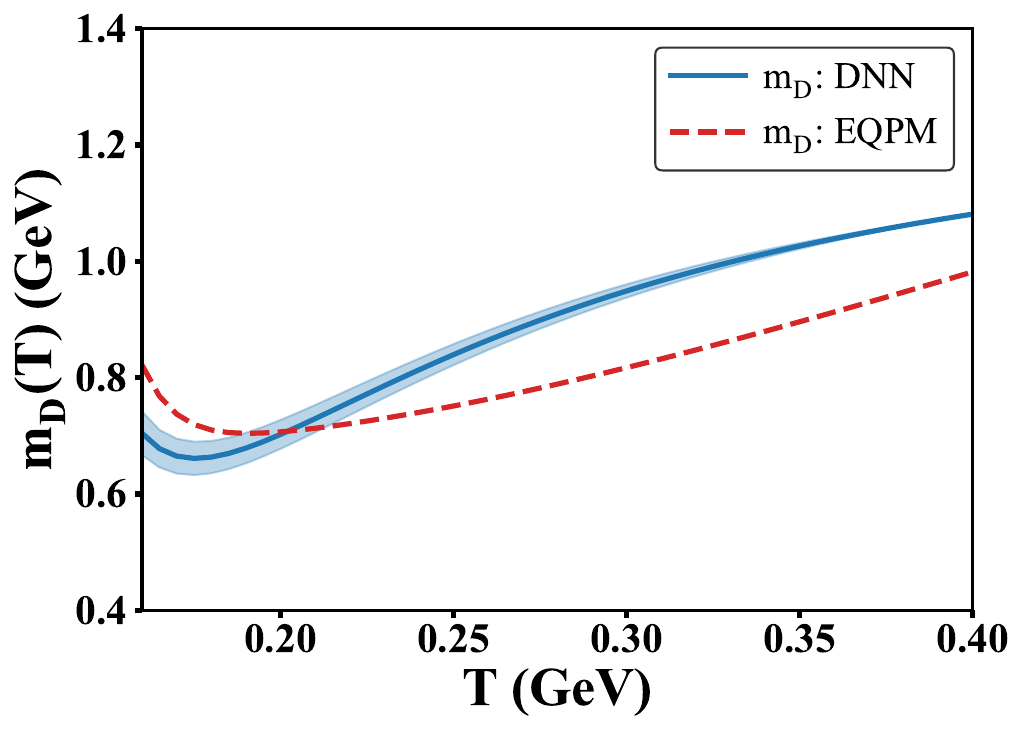}
    \caption{The screening mass as a function of temperature obtained from our DNN and EQPM. See the text for full details.
    }
    \label{fig:md}
\end{figure}

\begin{figure*}[ht!]
    \centering
    \includegraphics[width=0.48\textwidth]{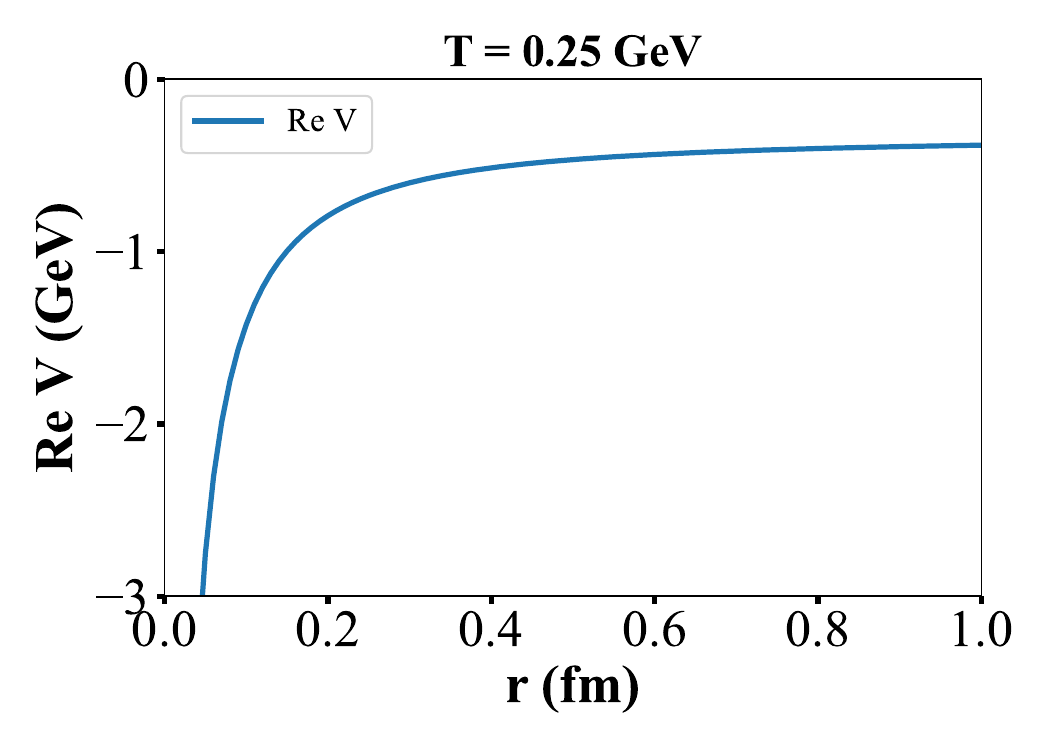}
    \includegraphics[width=0.48\textwidth]{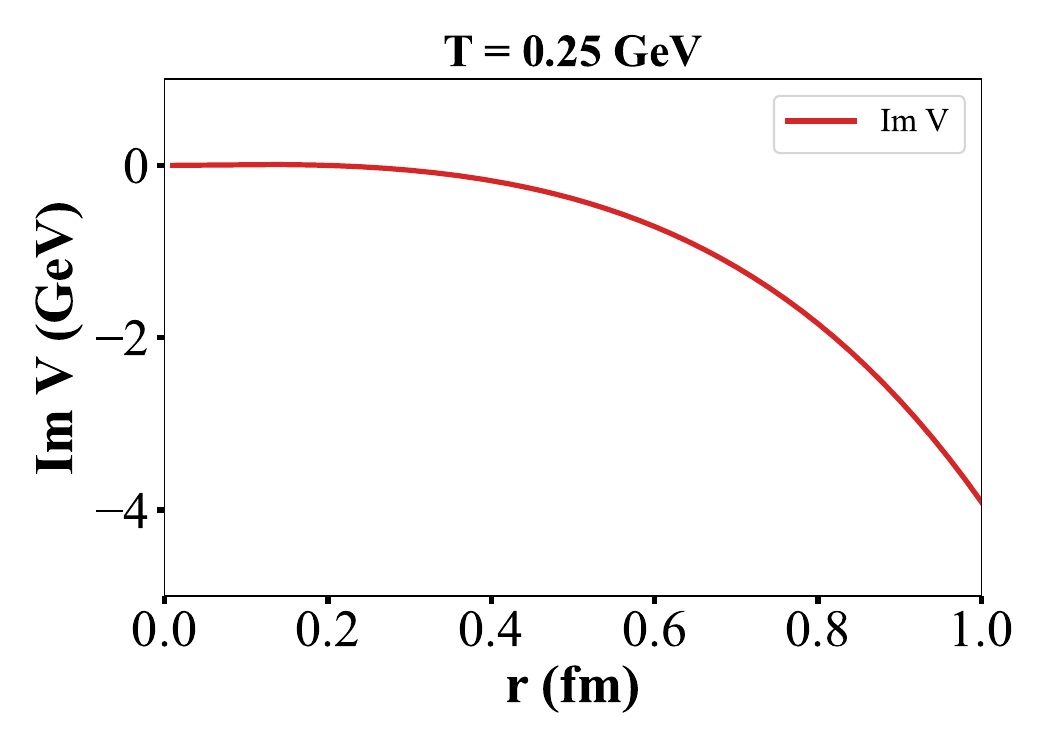}
    \caption{Left: Real part of the heavy quark potential as a function of inter-quark separation $r$ { at fixed temperatures, T=0.25 GeV}. Right: Corresponding imaginary part of the potential vs.\ $r$, capturing Landau damping and medium-induced broadening effects.
    }
    \label{fig:Potential_vs_r}
\end{figure*}
\begin{figure*}[ht!]
    \centering
    \includegraphics[width=0.48\textwidth]{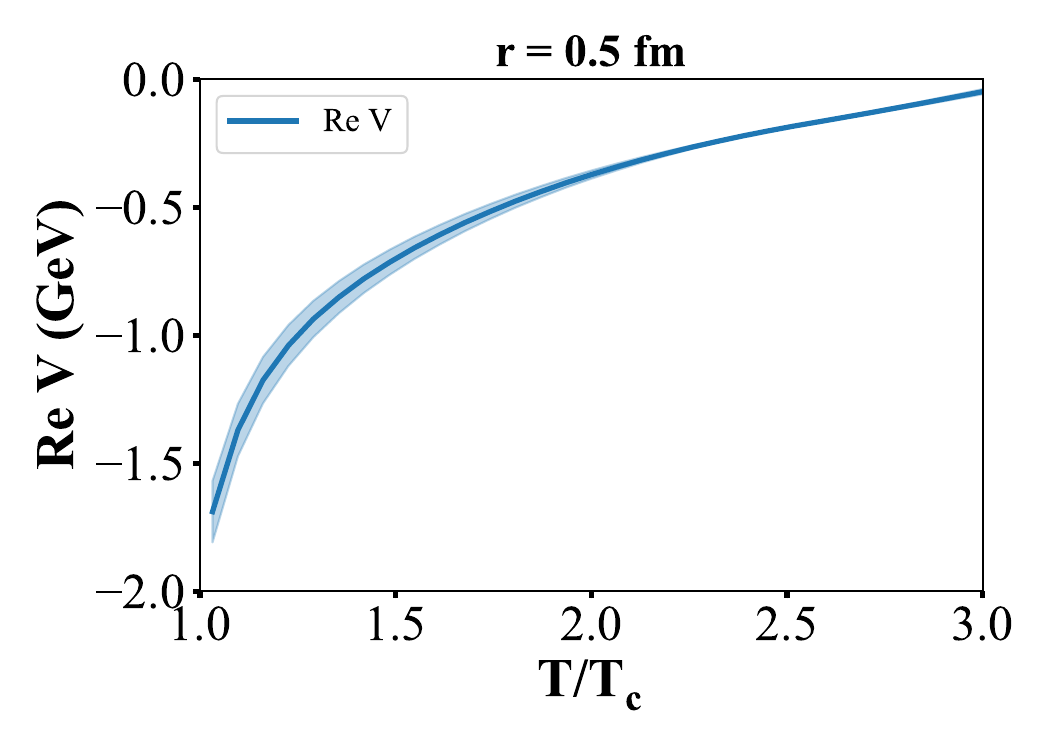}
    \includegraphics[width=0.48\textwidth]{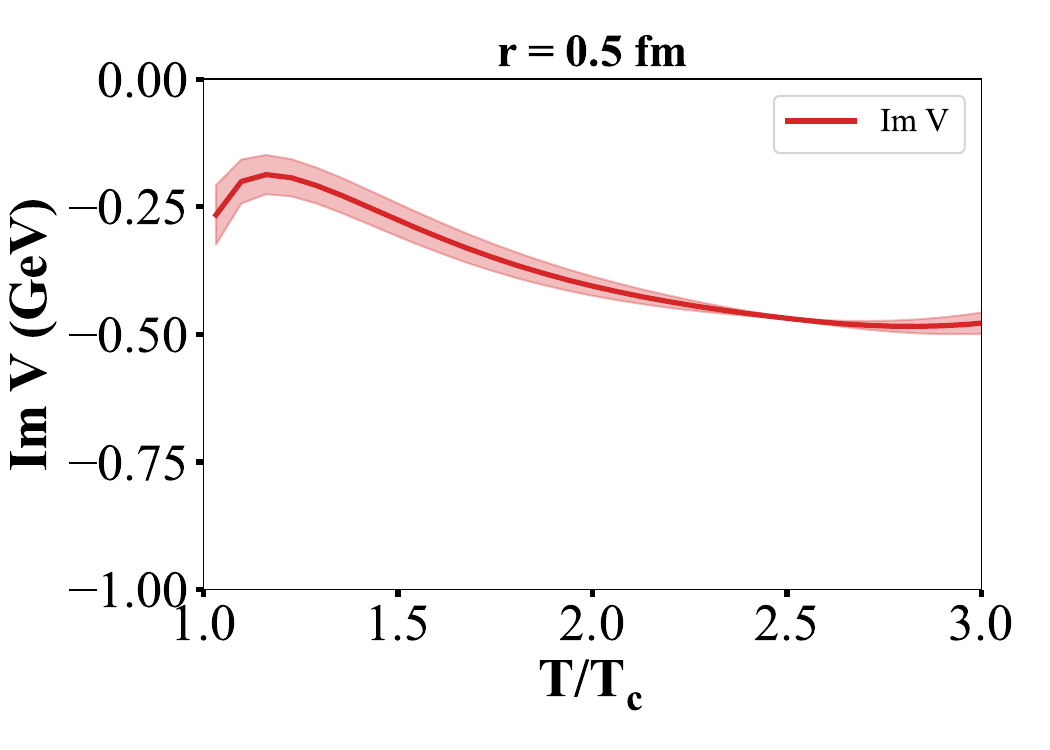}
    \caption{Temperature dependence of the heavy quark potential at a fixed inter-quark distance, { r=0.5 fm.} Left: Real part showing the gradual weakening of binding with increasing temperature. Right: Imaginary part illustrating the growth of thermal broadening as the QGP becomes hotter.
    }
    \label{fig:Potential_vs_T}
\end{figure*}

\begin{figure}[ht!]
    \centering
    \includegraphics[width=0.48\textwidth]{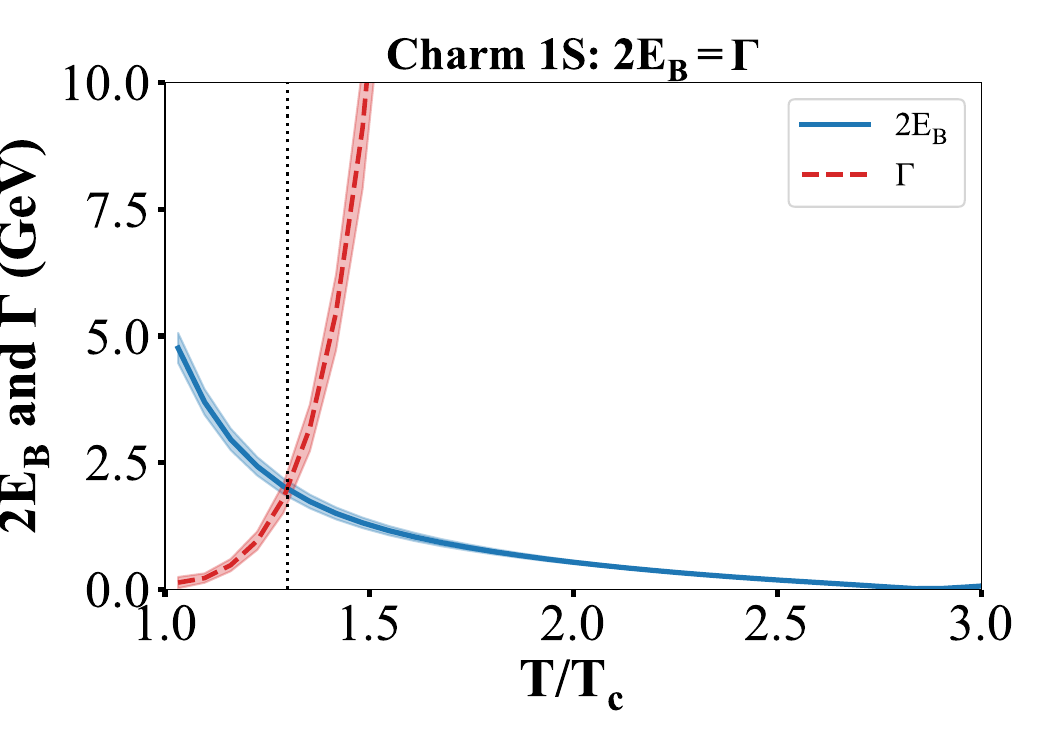}
    \caption{
    Binding energy and thermal width for Charmonium 1S-states ($J/\psi$) as a function of temperature. The dissociation temperature is obtained using the $2E_B = \Gamma$ criterion.
    }
    \label{fig:Criterion1_Charm}
\end{figure}

\begin{figure}[ht!]
    \centering
    \includegraphics[width=0.48\textwidth]{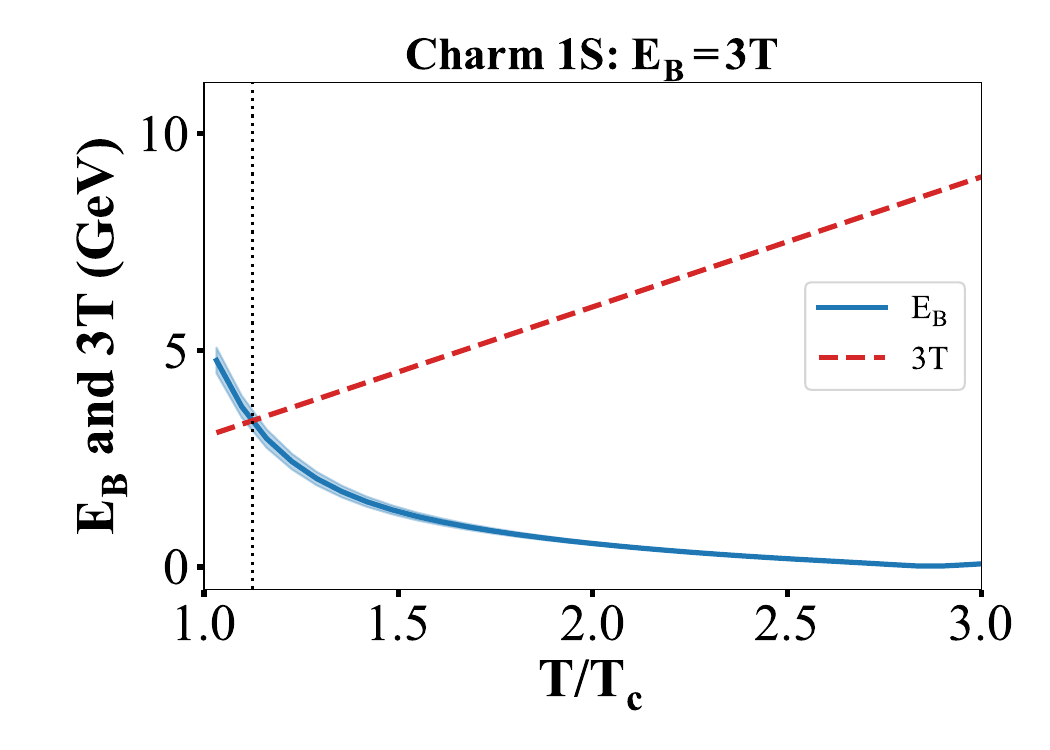}
    \caption{
    Binding energy for Charmonium 1S-states ($J/\psi$) as a function of temperature. The dissociation temperature is obtained using the lower bound criterion $E_B = 3T$.
    }
    \label{fig:Criterion2_Charm}
\end{figure}

\begin{figure*}[ht!]
    \centering
    \includegraphics[width=0.48\textwidth]{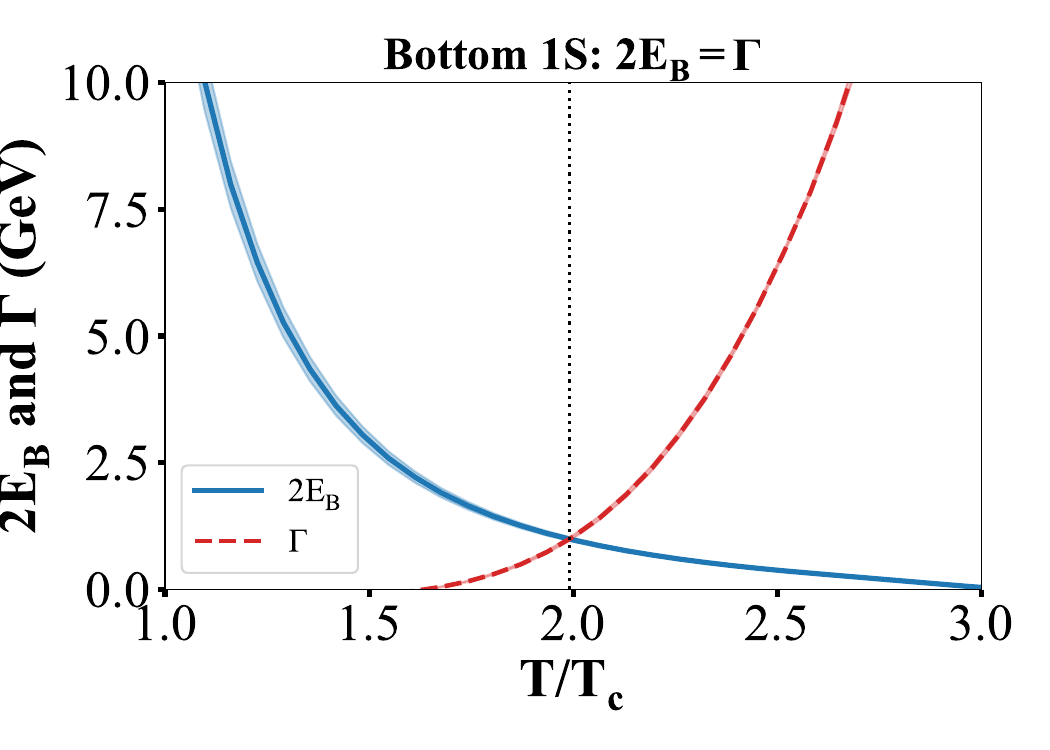}
    \includegraphics[width=0.48\textwidth]{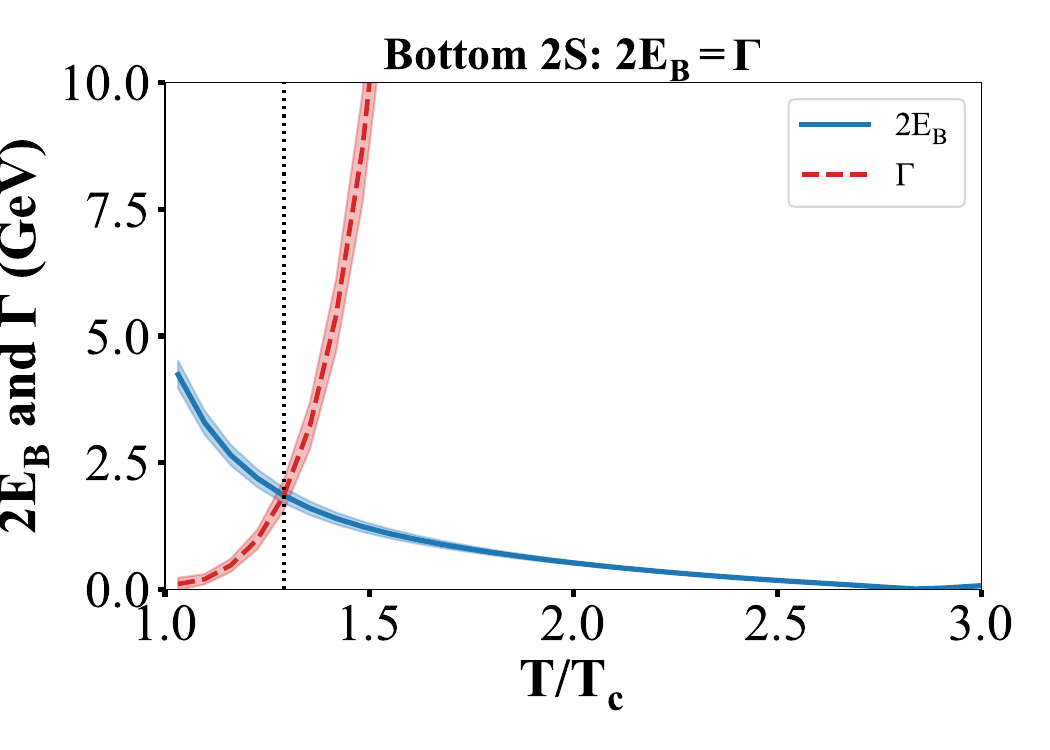}
    \caption{
    Binding energy and thermal width for Bottomonium states as a function of temperature. The dissociation temperature is obtained using the $2E_B = \Gamma$ criterion. The left panel shows the $\Upsilon$ ($1S$) state, while the right panel shows the $\Upsilon(2S)$ or $\Upsilon'$ state.
    }
    \label{fig:Criterion1_bottom}
\end{figure*}

\begin{figure*}[ht!]
    \centering
    \includegraphics[width=0.48\textwidth]{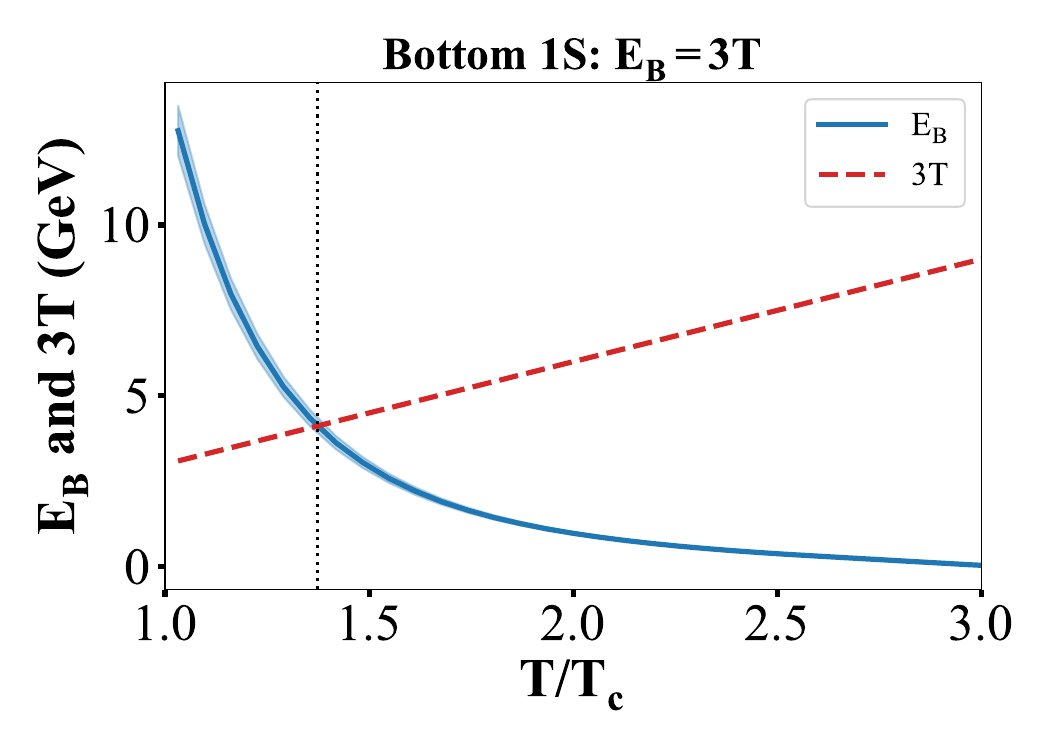}
    \includegraphics[width=0.48\textwidth]{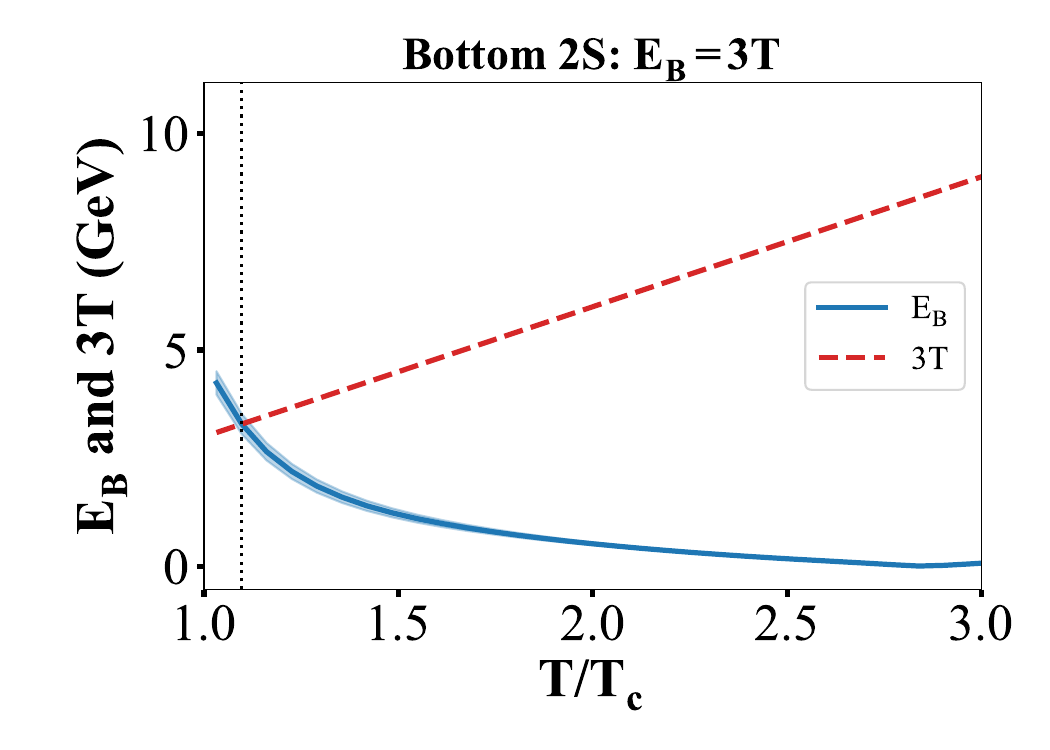}
    \caption{
    Binding energy for Bottomonium states as a function of temperature. The dissociation temperature is obtained using the lower bound criterion $E_B = 3T$. The left panel shows the $\Upsilon$ ($1S$) state, and the right panel shows the $\Upsilon(2S)$ or $\Upsilon'$ state.
    }
    \label{fig:Criterion2_bottom}
\end{figure*}

\begin{table*}[ht!]
\centering
\renewcommand{\arraystretch}{1.3}
\setlength{\tabcolsep}{10pt}
\begin{tabular}{ |p{3.8cm}||p{1.5cm}|p{1.5cm}|p{1.5cm}|p{1.5cm}|}
\hline
\multicolumn{1}{|c||}{Method / Reference} &
\multicolumn{4}{c|}{Dissociation Temperatures $T_d$ (in units of $T_c$)} \\
\cline{2-5}
\multicolumn{1}{|c||}{} &
$\Upsilon(1S)$ & $\Upsilon(2S)$ & $J/\psi$ & $\psi(2S)$ \\
\hline \hline
This work: ($2E_B = \Gamma$) & 1.99 & 1.29 & 1.30 & $\leq 1.00$ \\ 
This work: ($E_B = 3T$) & 1.38 & 1.10 & 1.13 & $\leq 1.00$ \\ 
\hline
\hline
Mocsy and Petreczky~\cite{Mocsy:2007jz} & 2.00 & 1.20 & 1.20 & $\!\leq\!1.00$ \\ 
\hline
Digal et al.~\cite{Digal:2001ue} & 2.31 & 1.10 & 1.10 & $\!<\!1.00$ \\ 
\hline
Satz~\cite{Satz:2005hx} & $>$4.00 & 1.60 & 2.10 & 1.12 \\ 
\hline
Blaschke et al.~\cite{Blaschke:2005jg} & 2.25 & 1.05 & 1.20 & $\!<\!1.00$ \\ 
\hline
Rethika et al.~\cite{Rethika:2021tdh} & 0.77 & 0.82 & 1.47 & 1.62 \\ 
\hline
Meng et al.~\cite{Meng:2018fco} & 5.81 & 1.56 & 2.06 & 1.13 \\ 
\hline
Jamal et al.~\cite{Jamal:2018mog} & 2.96 & 1.47 & 1.52 & $\!<\!1.00$ \\ 
\hline
Agotiya et al.~\cite{Agotiya:2016bqr} & 2.60 & 2.10 & 1.90 & 1.70 \\ 
\hline
\hline
\end{tabular}
\caption{
Comparison of the dissociation temperatures $T_d$ for selected quarkonium states. The first two rows show the results of this work using the thermal width criterion ($2E_B = \Gamma$) and the lower bound criterion ($E_B = 3T$). The lower rows present representative theoretical predictions from the literature~\cite{Mocsy:2007jz, Digal:2001ue, Satz:2005hx, Blaschke:2005jg, Rethika:2021tdh, Meng:2018fco, Jamal:2018mog, Agotiya:2016bqr}. All temperatures are in units of the QCD critical temperature $T_c$.}
\label{tab:qq_comparison}
\end{table*}

\section{Results and Discussion}
\label{sec:results}
We now present our results for the dissociation temperatures of various heavy quarkonia states within the QGP, based on an ML framework which combines lattice-QCD-informed Debye screening masses and running coupling $\alpha_s(T)$ with a robust potential model. Fig.~\ref{fig:md} displays the temperature dependence of the Debye mass extracted via our DNN training. { We trained the DNN 10 times and propagated the variance from \(m_{D}(T)\) through to the observables.} The smooth rise with increasing temperature ensures a consistent input to the in-medium potential.  {We have compared our results for the screening mass with predictions from the Effective Fugacity Quasi-Particle Model (EQPM). In this model, non-ideal medium effects are incorporated via a temperature-dependent fugacity. Our findings show a consistent agreement between the two. For further details on the EQPM, we refer the reader to Refs. \cite{Kumar:2017bja, Jamal:2020hpy, Jamal:2022ztl, Jamal:2017dqs}.}The behavior of the real and imaginary parts of the potential is shown in Figs.~\ref{fig:Potential_vs_r} and~\ref{fig:Potential_vs_T}. The real part demonstrates the expected screening: at larger separations or higher temperatures, the binding weakens significantly, reflecting color screening in the deconfined medium~\cite{Digal:2001ue,Mocsy:2007jz}. Meanwhile, the magnitude of the imaginary part grows monotonically with both inter-quark distance and temperature, encoding Landau damping and quark-antiquark scattering off thermal gluons~\cite{Laine:2006ns}.  

We solve the radial Schr\"odinger equation with this ML-enhanced complex potential to extract the temperature-dependent binding energy $E_B(T)$ and thermal width $\Gamma(T)$ for each quarkonium state. Figs.~\ref{fig:Criterion1_Charm} to ~\ref{fig:Criterion2_bottom} present the detailed evolution of these quantities for representative charmonium and bottomonium states under both dissociation conditions.
In the case of the width criterion, the curves for $2E_B$ and $\Gamma$ are shown together as functions of $T/T_c$. Their intersection indicates the temperature above which the thermal fluctuations are strong enough to overcome the binding within the lifetime of the resonance, signaling effective melting~\cite{Mocsy:2007jz}. This behavior is clearly visible in Figs.~\ref{fig:Criterion1_Charm} and~\ref{fig:Criterion1_bottom}: for each state, the crossing point shifts to lower temperatures for weaker binding, reflecting the sequential melting hierarchy.\
On the other hand, the lower-bound criterion shown in Figs.~\ref{fig:Criterion2_Charm} and~\ref{fig:Criterion2_bottom} compares the binding energy to the average thermal energy of the medium, approximated by $3T$~\cite{Agotiya:2008ie}. Once $E_B$ drops below this threshold, the bound state is unlikely to survive collisions with the surrounding deconfined medium. This provides a conservative estimate for the minimum temperature at which a given quarkonium state can persist. Finally, Table~\ref{tab:qq_comparison} consolidates the resulting dissociation temperatures obtained by both criteria for each quarkonium state and compares them to a range of benchmark predictions from other potential models, lattice extractions, and effective field theory analyses~\cite{Satz:2005hx,Blaschke:2005jg,Agotiya:2016bqr,Meng:2018fco,Jamal:2018mog}. The combined width-based upper-bound and lower-bound thermal estimate provides a credible bracket for the melting point of each state, capturing the known experimental hierarchy of suppression observed in heavy-ion collisions at RHIC and LHC.

For the tightly bound $\Upsilon(1S)$ ground state, our model predicts a dissociation temperature range of approximately $T_d \in [1.38, 1.99]\,T_c$, bounded from below by the $E_B = 3T$ criterion and from above by the $\Gamma = 2E_B$ width condition. This substantial survival window reflects the deeply bound nature of the state, which resists color screening and thermal broadening even at temperatures well above the deconfinement threshold.  This is consistent with CMS measurements that show a suppressed but non-zero nuclear modification factor $R_{AA} \approx 0.5$ for $\Upsilon(1S)$ even in the most central(0--10\%) Pb--Pb collisions at $\sqrt{s_{NN}} = 5.02$~TeV.~\cite{CMS:2012aa}. In contrast, the excited bottomonium state $\Upsilon(2S)$ exhibits a much lower dissociation threshold, with $T_d \in [1.10, 1.29]\,T_c$. This relatively narrow survival window implies that $\Upsilon(2S)$ mesons are significantly more vulnerable to medium effects, owing to their weaker binding and larger spatial extent. This theoretical result coheres with LHC data from CMS~\cite{CMS:2012aa, CMS:2018zza}, which show a dramatic suppression of $\Upsilon(2S)$ production, with $R_{AA}(\Upsilon(2S)) \approx 0.12$, nearly an order of magnitude lower than that of $\Upsilon(1S)$. Such differential suppression between the ground and excited states provides one of the clearest experimental signatures of sequential quarkonium melting in a thermalized QGP, affirming the predictive strength of our potential-based model enhanced with machine-learned screening dynamics.

For charmonium, our calculations yield a dissociation temperature range for the ground state $J/\psi$ of approximately $T_d \in [1.13, 1.30]\,T_c$. This moderate survival window reflects the fact that, while the $J/\psi$ is more tightly bound than its excited counterparts, its binding energy is still significantly smaller than that of $\Upsilon(1S)$, rendering it more sensitive to color screening and collisional broadening. This theoretical range aligns well with spectral function results from lattice QCD studies~\cite{Mocsy:2007jz} and recent phenomenological analyses that include non-perturbative corrections.
Experimentally, the ALICE and CMS collaborations have reported a clear but incomplete suppression of the $J/\psi$ yield in heavy-ion collisions. Specifically, the nuclear modification factor $R_{AA}(J/\psi)$ remains at about 0.6-0.7 for semi-central Pb--Pb events at the LHC~\cite{ALICE:2016flj}, indicating that a non-negligible fraction of $J/\psi$ states persist through the QGP phase. Moreover, the measured elliptic flow $v_2$ of $J/\psi$ mesons is finite, typically around 0.05-0.1~ \cite{CMS:2016mah}, supporting the idea that some fraction of $J/\psi$ production originates from regeneration via recombination of deconfined charm quarks in the late stages of QGP evolution~\cite{ALICE:2016flj}. Our calculated survival window, therefore, naturally accommodates both the direct production of primordial $J/\psi$ that survive color screening and the regenerated component that forms near hadronization.

The excited charmonium state $\psi(2S)$, however, shows a much weaker binding and thus a very limited survival probability in the QGP. { In our framework, both dissociation criteria predict $T_d$ near or below the critical temperature, $T_c$, confirming that even mild thermal fluctuations can completely screen its binding potential. Since our framework is not applicable to the confined region, we can not comment on that.}  However, this is consistent with the striking experimental observation by the CMS collaboration~\cite{CMS:2016psi2S} that the $\psi(2S)$ is significantly more suppressed than the $J/\psi$ in Pb--Pb collisions at $\sqrt{s_{NN}} = 5.02$~TeV, with an $R_{AA}$ well below 0.2 in central collisions. The enhanced suppression relative to the ground state underscores the significance of the binding energy hierarchy and the medium-induced imaginary potential in explaining the sequential melting pattern.

Taken together, our predicted hierarchy based on the ranges reads:
\[
T_d(\Upsilon(1S)) > T_d(J/\psi) \sim T_d(\Upsilon(2S)) > T_d(\psi(2S)),
\]
which aligns with the sequential suppression pattern observed in $R_{AA}$ measurements at RHIC and LHC~\cite{CMS:2012aa,ALICE:2016flj} 
\[
R_{AA}(\Upsilon(1S)) > R_{AA}(J/\psi) \sim R_{AA}(\Upsilon(2S)) > R_{AA}(\psi(2S)),
\]
and illustrate how the dissociation range captured by the dual criteria ($\Gamma = 2E_B$ and $E_B = 3T$) constrains the expected suppression hierarchy and its temperature dependence. Notably, our ML-enhanced potential tightens the theoretical uncertainties by providing a non-perturbative screening mass, thus bridging lattice QCD constraints with continuum phenomenology. We present a comparison of our results at different grid sizes in Appendix \ref{sec:FDM1} to verify numerical stability. Overall, these results support the interpretation of quarkonia as sensitive QGP thermometers. The approach naturally provide an improved baseline for interpreting quarkonium suppression at RHIC and LHC, reproduces the observed trends in suppression and flow observables, and offers a systematic way to extend to P-wave quarkonia, finite-momentum effects, and eventually to open quantum system treatments that can predict time-resolved $R_{AA}$ and $v_2$ in dynamic, evolving media~\cite{Brambilla:2017zei}. Such developments will further solidify heavy quarkonia as precision probes for the emergent collective properties of the QGP.

\section{Summary and Conclusions}
\label{sec:Con}
In this work, we have developed a robust, data-driven framework to investigate the in-medium behavior and dissociation characteristics of heavy quarkonia in the QGP. To achieve this, we employed an ML approach using a DNN trained on lattice QCD data to obtain a reliable, non-perturbative estimate of the Debye screening mass $m_D(T)$ and the strong coupling constant $\alpha_s(T)$. These ML-improved quantities were then used to construct a medium-modified Cornell potential that accurately incorporates both color screening and dynamical Landau damping effects through a complex-valued potential. Solving the radial Schr\"odinger equation with this in-medium potential, we extracted the temperature-dependent binding energies and thermal widths for different quarkonium states, $J/\psi$, $\psi(2S)$, $\Upsilon(1S)$, and $\Upsilon(2S)$. The thermal width, derived as a first-order perturbative estimate from the imaginary part of the potential, quantifies the decay rate due to medium-induced interactions such as gluodissociation and scattering with thermal gluons.

To determine the dissociation temperatures, we implemented two complementary physical criteria. The first, the conventional width criterion ($2E_B = \Gamma$), identifies the point at which the resonance peak broadens sufficiently to lose its bound state character, considered as an upper bound \cite{Mocsy:2007jz}. The second, the lower bound criterion ($E_B = 3T$), asserts that a quarkonium state cannot survive once its binding energy falls below the scale of typical thermal fluctuations in the medium \cite{Agotiya:2008ie}. This dual-criterion approach provides a more constrained and physically motivated estimate of the dissociation temperature range for each state. Our results suggest that the ground-state bottomonium \(\Upsilon(1S)\) may remain intact up to approximately \(1.99\,T_c\), reinforcing its status as a reliable probe of the hottest phases of the QGP. In contrast, the excited bottomonium state \(\Upsilon(2S)\) is expected to survive only up to a considerably lower temperature near \(1.29\,T_c\). For charmonia, the \(J/\psi\) is predicted to dissolve around \(1.3\,T_c\), whereas the more weakly bound \(\psi(2S)\) state melts close to the critical temperature. These findings are consistent with the sequential suppression patterns observed experimentally in heavy-ion collisions at RHIC and the LHC.

A detailed comparison with various theoretical estimates from the literature demonstrates that our ML-augmented framework yields dissociation temperatures in good agreement with lattice QCD constraints and modern potential model predictions, while also reducing dependence on purely perturbative approximations. Overall, this study highlights the utility of integrating ML with potential model analyses to bridge non-perturbative lattice data and phenomenological quarkonium observables. 

\subsection{Future Prospects}
\label{Future}

While the present study provides a comprehensive estimate of quarkonium dissociation temperatures using an ML-enhanced potential model, several avenues remain open for further refinement and deeper insights. First, an immediate extension of this work will involve solving the full time-independent Schr\"odinger equation with the potential itself modeled directly by a deep neural network. By training the DNN to learn the entire complex-valued in-medium potential from lattice QCD correlators or spectral functions, one can bypass analytic ansatze and capture more subtle non-perturbative effects beyond the conventional dielectric model. This data-driven approach promises a more flexible and accurate description of quarkonium binding and broadening at all temperatures.

Second, an important improvement will be to generalize the current static potential framework to an open quantum system treatment. In such a formalism, the heavy quarkonium is treated as an evolving subsystem interacting with a thermalized QGP environment, described dynamically via stochastic Langevin or Lindblad equations. This would allow for a first-principles computation of real-time decoherence, non-Markovian effects, and medium-induced transitions between bound and unbound states, providing a more realistic picture of quarkonium survival and regeneration in heavy-ion collisions.

In addition to these methodological advancements, future studies will systematically extend the present analysis to include P-wave states, explore finite momentum effects, and incorporate dynamical quark masses with explicit chiral symmetry restoration. Such improvements will further strengthen the connection between theory and the wealth of high-precision quarkonium data from RHIC, LHC, and forthcoming facilities, such as the Electron-Ion Collider.

\appendix

\section{DNN details} \label{sec:DNN}

\begin{figure*}[ht!]
    \centering
    \includegraphics[width=1.0\linewidth]{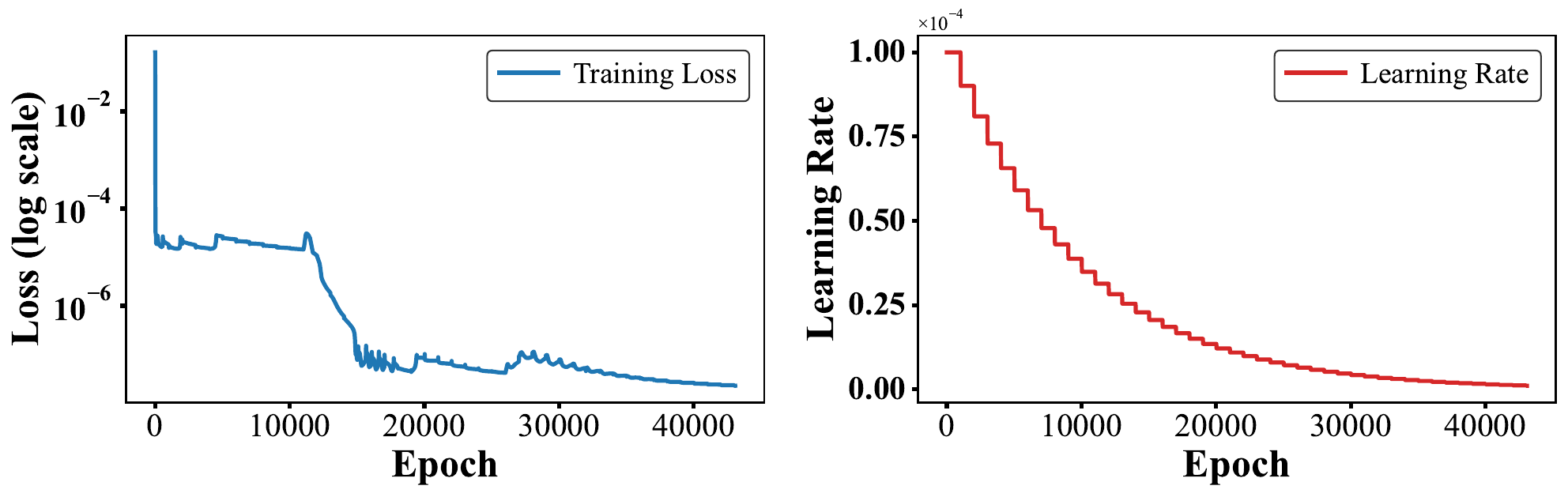}
    \caption{Training loss (upper panel) and learning rate (lower panel) as functions of epoch}
    \label{fig:train}
\end{figure*}

In Fig.~\ref{fig:train} we display the training diagnostics of the DLQPM, namely the total loss and the learning rate as functions of the training epoch. The loss decreases monotonically and saturates after approximately \(5\times 10^{4}\) epochs, indicating stable convergence of the optimisation, while the learning-rate schedule ensures a smooth approach to the minimum. This behaviour is consistent across the independent trainings used for the uncertainty analysis described in Sec.~\ref{sec:ML}.

\section{Numerical Performance of FDM} \label{sec:FDM}

This section outlines the finite-difference method (FDM) employed in our study and compares it with other numerical approaches. In addition to standard benchmarks, we include a direct comparison of different grid sizes within our current results, illustrating both the accuracy and convergence properties of the solver.

\subsection{Numerical method (we apply)}
We solve the $s$-wave radial Schrödinger equation on a uniform grid in coordinate space. The radial coordinate is discretized between $r_{\min}$ and $r_{\max}$ with $N$ points, and the second derivative is approximated by a centered finite-difference scheme. This leads to a real, symmetric, tridiagonal Hamiltonian matrix, subject to Dirichlet boundary conditions at the endpoints of the grid~\cite{Press:2007nr,Thijssen:2007cp,LeVeque:2007fd,Ascher:1995bvp}. The lowest eigenvalues and eigenfunctions are then obtained with an iterative sparse-matrix eigensolver, which is well suited for large tridiagonal systems~\cite{Lehoucq:1998arpack}.

The real part of the in-medium potential enters directly in the Hamiltonian, together with the centrifugal contribution for higher orbital angular momenta (though we focus here on $l=0$ states). The imaginary part of the potential is used afterwards to estimate thermal widths from the normalized wavefunctions. To control numerical instabilities near $r=0$, a small-distance regulator is applied when evaluating the kernels.

The discretization error of this scheme decreases quadratically with the grid spacing, so energy levels converge as the inverse square of the number of grid points. In practice, we vary the resolution $N$ while keeping the domain $[r_{\min},r_{\max}]$ fixed, and check that the dissociation observables are stable. Unless otherwise noted, calculations use $r_{\min}=0.01\ \mathrm{fm}$, $r_{\max}=30.01\ \mathrm{fm}$, and a baseline resolution of $N=4000$ grid points, with additional runs at coarser and finer grids to verify convergence.

\subsection{Benchmarks with other potentials having analytical spectra.}

To validate the solver and its units, we first consider simple potentials with known spectra. The harmonic oscillator (HO) and Coulomb problems serve as analytic benchmarks, while a Woods--Saxon (WS) well provides a more realistic test case without closed-form solutions. The corresponding potentials are given below:

\begin{equation}
V_{\mathrm{HO}}(r)=\tfrac12\,\mu\,\omega^2\,(r/\hbar c)^2,\qquad
E_0^{\mathrm{HO}}=\tfrac32\,\omega,
\end{equation}
\begin{equation}
V_{\mathrm{C}}(r)=-\alpha\,\frac{\hbar c}{r},\qquad
E_0^{\mathrm{C}}=-\frac{\mu\,\alpha^2}{2},
\end{equation}
\begin{equation}
V_{\mathrm{WS}}(r)=-\frac{V_0}{1+\exp[(r-R)/a]}\quad(\text{no closed form}),
\end{equation}

The corresponding potentials and ground-state energies are listed in Table~\ref{tab:fdm_bench}. In all cases, the finite-difference solver reproduces the analytic results to very high accuracy, confirming both the discretization scheme and our unit conventions. The Woods--Saxon problem illustrates the stability of the method even when exact energies are unavailable.

\begin{table*}
\centering
\renewcommand{\arraystretch}{1.3}
\setlength{\tabcolsep}{10pt}
\begin{tabular}{|l|p{3.0cm}|p{1.0cm}|p{1.20cm}|p{2.0cm}|p{2.0cm}|p{1.8cm}|}
\hline
\multicolumn{1}{|c|}{Potential} &
\multicolumn{1}{c|}{Parameters} &
\multicolumn{1}{c|}{$N$} &
\multicolumn{1}{c|}{$r_{\max}$ [fm]} &
\multicolumn{2}{c|}{Ground-state energy [GeV]} &
\multicolumn{1}{c|}{Rel.\ error} \\
\cline{5-6}
\multicolumn{1}{|c|}{} &
\multicolumn{1}{|c|}{} &
\multicolumn{1}{|c|}{} &
\multicolumn{1}{|c|}{} &
\textbf{Numerical} & \textbf{Exact} &
\multicolumn{1}{c|}{} \\
\hline \hline

Harmonic oscillator & \(\omega=0.30\) & 2000 & 15.0 & 0.449986 & 0.450000 & \(3.2\times10^{-5}\) \\
\hline
Coulomb (vacuum)    & \(\alpha=0.40\) & 2000 & 15.0 & \(-0.059998\) & \(-0.060000\) & \(3.9\times10^{-5}\) \\
\hline
Woods--Saxon        & \(V_0{=}0.50,\,R{=}2.0,\,a{=}0.5\) & 2000 & 15.0 & \(-0.378654\) & --- & --- \\
\hline
\end{tabular}
\caption{Benchmark ground-state energies computed with the finite-difference solver and comparison to exact values when available.}
\label{tab:fdm_bench}
\end{table*}

\begin{figure}[ht!]
\centering
\includegraphics[width=0.9\linewidth]{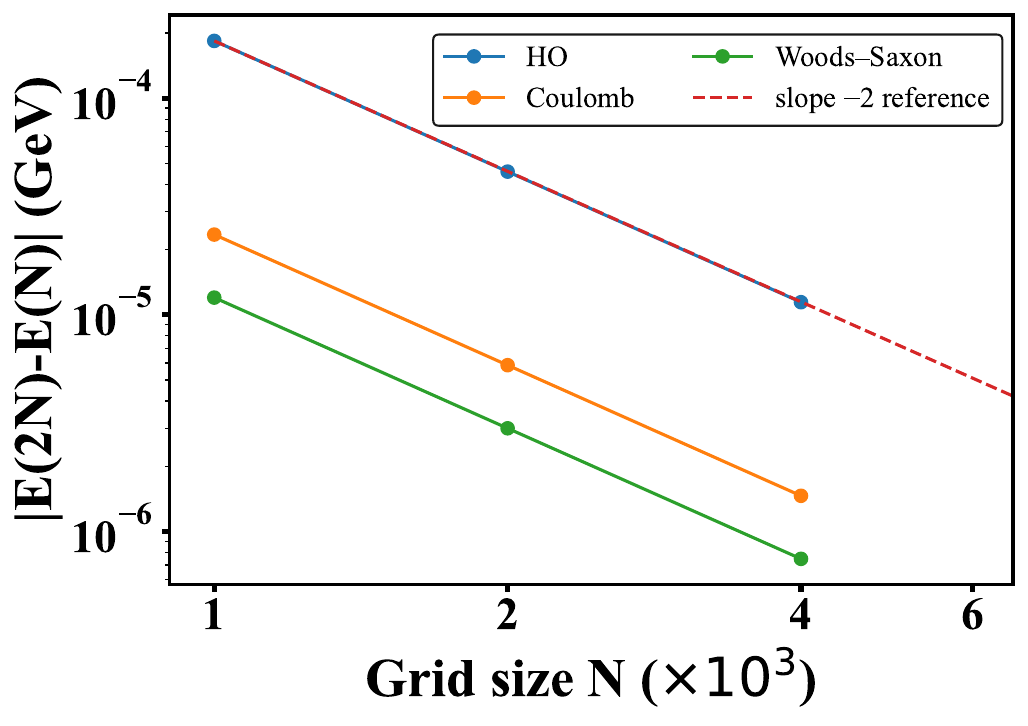}
\caption{Grid-refinement residuals \(|E(2N)-E(N)|\) as a function of \(N\) (log-log scale). The dashed line indicates a reference slope of \(-2\). The three series (HO, Coulomb, WS) follow the \(N^{-2}\) trend.}
\label{fig:fdm_convergence_plot}
\end{figure}

\subsection{Convergence with grid size \(N\).}

We next assess numerical convergence by refining the uniform grid in a fixed spatial domain. As expected for a second-order stencil, the errors decrease quadratically with \(1/N\). Table~\ref{tab:fdm_conv} summarizes the ground-state energies at successive refinements, while Fig.~\ref{fig:fdm_convergence_plot} shows the log–log behavior of residuals. The observed slopes closely follow the theoretical \(-2\) trend, and Richardson extrapolation provides values indistinguishable from the exact results. The Woods--Saxon benchmark likewise exhibits rapid stabilization, with negligible change once \(N\) exceeds a few thousand.

\begin{table*}
\centering
\renewcommand{\arraystretch}{1.3}
\setlength{\tabcolsep}{10pt}
\begin{tabular}{|l|p{2.1cm}|p{2.1cm}|p{2.1cm}|p{2.1cm}|}
\hline
\multicolumn{1}{|c||}{Potential} &
\multicolumn{4}{c|}{Ground-state energy $E_0$ [GeV] vs.\ grid size $N$} \\
\cline{2-5}
\multicolumn{1}{|c||}{} &
$N{=}500$ & $N{=}1000$ & $N{=}2000$ & $N{=}4000$ \\
\hline
\hline 
Harmonic oscillator & 0.449756 & 0.449940 & 0.449986 & 0.449997 \\
\hline
Coulomb (vacuum)    & \(-0.059968\) & \(-0.059992\) & \(-0.059998\) & \(-0.059999\) \\
\hline
Woods-Saxon        & \(-0.378669\) & \(-0.378657\) & \(-0.378654\) & \(-0.378653\) \\
\hline
\end{tabular}
\caption{Grid-refinement study at fixed \(r_{\max}=15~\mathrm{fm}\). The observed slopes of \(|E(2N)-E(N)|\) vs.\ \(N\) are close to \(-2\), consistent with the second-order stencil. Richardson extrapolations and internal error bars are quoted in the text.}
\label{tab:fdm_conv}
\end{table*}

\subsection{Reporting practice for in-medium potentials.}

For realistic in-medium heavy-quark potentials, no analytic reference exists. In these cases, we accompany quoted binding energies with systematic convergence checks: (i) refinement in the number of grid points to confirm the expected \(N^{-2}\) scaling, and (ii) variation of the spatial cutoff \(r_{\max}\) to control finite-volume effects. Dirichlet boundary conditions are imposed consistently. With these safeguards, the residual numerical uncertainty is much smaller than the physical variation of the binding energy across the relevant temperature range.


\subsection{Convergence check with grid size $N$ of current numerical analysis.}\label{sec:FDM1}

\begin{table*}[tbh]
  \begin{center}
    \begin{tabular}{ |p{3.8cm}||p{1.5cm}|p{1.5cm}|p{1.5cm}|p{1.5cm}|p{1.5cm}|p{1.5cm}|p{1.5cm}|p{1.5cm}| }
      \hline
\hline
\multicolumn{1}{|c||}{State / Criterion} &
\multicolumn{8}{c|}{Dissociation temperatures $T_d/T_c$ ($T_c=0.15\,\mathrm{GeV}$) vs.\ grid size $N$} \\
\hline
\multicolumn{1}{|c||}{\scriptsize (lower row: $|\Delta|$ vs.\ $N{=}500$)} &
$N{=}500$ & $N{=}1000$ & $N{=}1500$ & $N{=}2000$ &
$N{=}3000$ & $N{=}4000$ & $N{=}5000$ & $N{=}6000$ \\
\hline

$\Upsilon(1S)$ \;$(2E_B=\Gamma)$
& 3.319 & 2.616 & 2.253 & 2.125 & 2.029 & 1.992 & 1.969 & 1.956 \\
$|\Delta|$ vs $N{=}500$
& 0.000 & 0.703 & 1.065 & 1.194 & 1.289 & 1.327 & 1.350 & 1.363 \\
      \hline

$\Upsilon(1S)$ \;$(E_B=3T)$
& 2.490 & 1.883 & 1.606 & 1.488 & 1.405 & 1.375 & 1.359 & 1.349 \\
$|\Delta|$ vs $N{=}500$
& 0.000 & 0.607 & 0.884 & 1.002 & 1.085 & 1.116 & 1.131 & 1.141 \\
      \hline\hline

$\Upsilon(2S)$ \;$(2E_B=\Gamma)$
& 1.813 & 1.641 & 1.485 & 1.392 & 1.317 & 1.291 & 1.274 & 1.264 \\
$|\Delta|$ vs $N{=}500$
& 0.000 & 0.171 & 0.327 & 0.421 & 0.496 & 0.521 & 0.539 & 0.549 \\
      \hline

$\Upsilon(2S)$ \;$(E_B=3T)$
& 1.286 & 1.254 & 1.196 & 1.150 & 1.113 & 1.097 & 1.092 & 1.087 \\
$|\Delta|$ vs $N{=}500$
& 0.000 & 0.033 & 0.091 & 0.136 & 0.174 & 0.189 & 0.194 & 0.199 \\
      \hline\hline

$J/\psi$ \;$(2E_B=\Gamma)$
& 2.332 & 1.573 & 1.415 & 1.359 & 1.317 & 1.299 & 1.291 & 1.284 \\
$|\Delta|$ vs $N{=}500$
& 0.000 & 0.758 & 0.917 & 0.972 & 1.015 & 1.033 & 1.040 & 1.048 \\
      \hline

$J/\psi$ \;$(E_B=3T)$
& 1.901 & 1.352 & 1.206 & 1.160 & 1.135 & 1.125 & 1.120 & 1.115 \\
$|\Delta|$ vs $N{=}500$
& 0.000 & 0.549 & 0.695 & 0.740 & 0.766 & 0.776 & 0.781 & 0.786 \\
      \hline
    \end{tabular}
\caption{Dissociation temperatures \(T_d/T_c\) vs grid size \(N\) for selected quarkonia.
Lower row in each cell shows the absolute difference relative to \(N{=}500\).}
    \label{tab:grid}
  \end{center}
\end{table*}

The objective of the analysis is to examine how the dissociation temperatures of selected quarkonia states vary with the numerical grid size used in the computation. The states of interest include \(J/\psi\) (charmonium ground state), \(\Upsilon(1S)\) (bottomonium ground state), and \(\Upsilon(2S)\) (bottomonium first excited state). For each state, two physical dissociation criteria are considered: 
(1) \(2E_B = \Gamma\), and 
(2) \(E_B = 3T\), where \(E_B\) denotes the binding energy and \(\Gamma\) the thermal width.

The dissociation temperatures are expressed in units of the critical temperature, \(T_d/T_c\), with \(T_c = 0.15\, \text{GeV}\). The benchmark grid chosen is \(N=500\), and the results from larger grids are compared against it. The absolute difference \(|\Delta|\) with respect to the benchmark is used as a measure of numerical convergence.

\begin{figure}[ht!]
\centering
\includegraphics[width=0.9\linewidth]{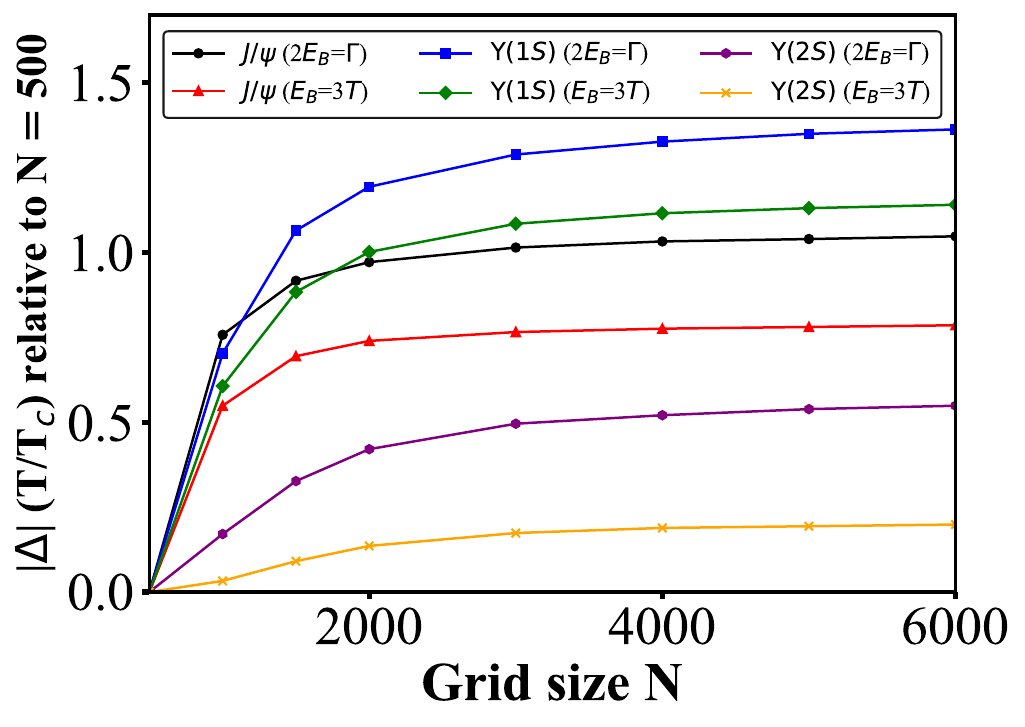}
\caption{Comparison of dissociation temperatures \(T_d/T_c\), with \(T_c = 0.15\, \text{GeV}\) for various quarkonia states and criteria  vs. grid size \(N\).}
\label{fig:grid_N}
\end{figure}

The results are shown in Table: \ref{tab:grid} which indicate that:
\begin{itemize}
\item For all states and criteria, the dissociation temperature decreases systematically as the grid size increases. This behavior reflects the fact that coarse grids tend to overestimate the binding, thus delaying the dissociation.
\item The difference \(|\Delta|\) grows substantially for finer grids, stabilizing for \(N \gtrsim 3000\). This demonstrates the need for large grid sizes to achieve accurate and stable dissociation predictions.
\item The ground state bottomonium, \(\Upsilon(1S)\), exhibits the largest differences with increasing grid size. This sensitivity highlights the stronger binding of bottom quarkonia and the numerical challenges in resolving their dissociation accurately.
\item The excited bottomonium state, \(\Upsilon(2S)\), shows smaller but still noticeable variations, consistent with its weaker binding and earlier dissociation.
\item The charmonium state \(J/\psi\) exhibits intermediate sensitivity, lying between the bottomonium ground and excited states.
\end{itemize}

Overall, the analysis confirms that grid resolution is important in numerical studies, specifically in the current study of dissociation temperature calculations. The Fig.\ref{fig:grid_N} shows that the benchmarks at small grid sizes (\(N=500\)) are insufficient for quantitative precision. Only at higher resolutions does the dissociation temperature converge to stable values, providing reliable physical insight into the quarkonia melting pattern in the quark-gluon plasma.

\section*{Data Availability}
 For further information about the data and code used in this study, please refer to the GitHub repository link provided ~\cite{DLQPM:FPLi}.

Data and code supporting this study are available upon reasonable request.

\section*{Acknowledgement}

This work was supported in part by Natural Science Foundation of China (NSFC) under grant Nos. 12225503 and 12435009, and by National Key Research and Development Program of China under Grant No. 2020YFE0202002. The work is also partly supported by the Cross Research Project of Fundamental Research Funds for Central Universities of Central China Normal University in 2025: "Advanced Detection and Artificial Intelligence at the Frontiers of Physics"(No.30101250317). We gratefully acknowledge the extensive computing resources provided by the Nuclear Science Computing Center at Central China Normal University (NSC$^3$). 

\bibliography{ref1}
\end{document}